\documentclass[11pt,en,authoryear]{elegantpaper}
\usepackage{supertabular}
\usepackage{setspace}
\usepackage{hyperref}
\hypersetup{hypertex=true, colorlinks=true,linkcolor=blue,urlcolor=blue,citecolor=blue,plainpages=false,breaklinks=true}
\usepackage{authblk}
\title{{\color{black} Natural resources balance sheets accounting: theoretical framework and practice in the Shaanxi province of China \footnote{This research is supported by the National Natural Science Foundation of China (Nos. 72203166), the China Postdoctoral Science Foundation (Nos. 2019M663756), and the Fundamental Research Funds for Central Universities (Nos. 2020XJS23).}}}
\author[1,2]{Wentao Wang\thanks{Corresponding author:wtwang\_cn@163.com}}
\author[2]{Guoping Li}
\author[3]{Andreas Kontoleon}
\author[2]{Yiming Ma}
\author[2]{Weishan Guo}
\affil[1]{School of Equipment Management and Unmanned Aerial Vehicle Engineering, Air Force Engineering University, Xi'an, Shaanxi, China}
\affil[2]{School of Economics and Finance, Xi'an Jiaotong University, Xi'an, Shaanxi, China}
\affil[3]{Department of Land Economy, University of Cambridge, Cambridge CB3 9EP, United Kingdom}

\date{}

\begin{document}
\maketitle
\begin{abstract}
To achieve sustainable development, there is widespread of the need to protect natural resource and improve government oversight in achieving China's economic security and ecological civilization. Compilation of natural resources balance sheet (NRBS) and enhancement of resources management are becoming an important topic in China. How to compile NRBS to affix the responsibility for government and officials for inadequate supervision is still not resolved satisfactorily. This paper proposes the NRBS to enable governments to identify the importance of natural resource restoration and to hold leading cadres accountable for a lack of adequate supervision. The NRBS consist of three accounts: natural resource assets, natural resource liabilities, and net worth. Important components of the NRBS for the liabilities account with a property rights regime are developed to measure and assign responsibility. The compilation of an NRBS is applied to the Chinese province of Shaanxi as an illustration to demonstrate that the accounting framework and the compilation steps are tractable using financial methods and available data. The accounting results of natural resource assets and liabilities unveil the threat to resource management and the policy implications to government and officials. Finally, the advantages and limitations of NRBS are discussed. \\

\end{abstract}

  \ \ \ \ \ \ \par

  \ \ \ \ \ \ \par

\section{Introduction}
Over the past quarter century, the global economy as measured by gross domestic product has approximately doubled in constant dollar terms. At the same time, this rapid economic growth has been accompanied by the massive consumption of natural resources \citep{wang2020}. Environmental destruction (e.g., cities besieged by waste, the decline in the water table, and suspended-particle emissions) has raised widespread concern about human health and is a major contributor to premature mortality \citep{srinivasan2008debt,tschofen2019fine}. The loss and degradation of ecosystem assets and services has raised serious concern about the resilience and sustainability of ecosystems \citep{ouyang2020using}. Moreover, the international community has tried and failed to find a path to slow the worldwide decline in biodiversity \citep{rounsevell2020biodiversity}. However, all of the above anthropogenic damage to nature has not been fully captured or adequately quantified in economic systems, resulting in such issues often being given too little weight in decision making.\par

Recognizing that economic security and sustainable development are seriously threatened by the destruction of the natural environment, the Chinese government has abandoned its blind pursuit of GDP growth and realized the importance of natural resource restoration and management. Since the 18th National Congress of Communist Party of China, the national government proposed establishing a natural resource balance sheets (NRBS) and designing a pilot scheme in selected regions \citep{zhiming2014natural}. As rightly pointed out by \cite{zhu2021natural}, how to investigate the principles and compilation of NRBS is essential for securing natural resource and constructing ecological civilization. Moreover, the compilation of NRBS has becoming an important tool to enhance the efficiency of government's oversight and resource management \cite{song2019}. Recently, the Natural Resources Balance Sheet Preparation System is promulgated by the National Bureau of Statistics in October 2020. This policy provides an vital support and evidence for the management of nature resources \citep{pan2022liability,chen2022developing}. The primary purpose of NRBS accounting is to enable the government and officials to identify the costs incurred by the entities responsible for preventing the degradation of nature and to hold leading cadres accountable for a lack of adequate supervision. The NRBSs are also used to assess the performance of the local governments in efficiency utilization of resources \citep{song2020influences}. However, it is difficult to construct the key definitions and relationships between natural resources assets and liabilities in compiling a regional NRBS \citep{shi2020research,opaluch2020liability}. Moreover, how to measure and assign responsibility of laking of adequate supervision by the government and officials is still not resolved satisfactorily \citep{song2019,yang2020}. Therefore, this paper proposes an accounting framework, including key accounting concepts and steps, for the compilation of a regional NRBS and a case study. This framework is used to identify the latent risk factors in natural resource liabilities and to ensure security in resource utilization.\par
The remainder of this paper is organized as follows. Section 2 presents the recent developments in the System of National Accounts (SNA), natural resource accounting (NRA), the System of Environmental Economic Accounting (SEEA), natural resource liabilities, and the NRBS. Section 3 describes the accounting framework, key accounting concepts and compilation steps for our approach. Section 4 presents a case study of Shaanxi Province, with detailed valuation methods and data sources for each item on the NRBS reported in the Supplemental Materials. Finally, Section 5 provides the policy implications for resource management.\par
\section{Literature Review}
\subsection{SNA, NRA, and SEEA}
As an international statistical standard, the SNA is used worldwide as a tool for assessing economic performance and measuring wellbeing and the progress made by economic units in formulating policy \citep{Allin2017From}. Since its inception, the lack of a production boundary has been a limitation in its ability to describe the linkages between economic and environmental systems \citep{Eigenraam2018}. As pointed out by \cite{LaNottea2020}, the overexploitation of natural resources and emissions of pollutants are typically included within the family of negative `externalities', which do not have a place within the SNA. Another problem with the concentration of analysis on accounting concepts is the lack of consideration given to ecological assets and services \citep{LaNottea2020}. Among the accounting indicators in the SNA, one of the most widely accepted measures of economic progress is GDP, which provides a snapshot of the GDP of a country. As pointed out by \cite{Hartwick1990Natural}, however, the method by which GDP is calculated may neglect the depletion of natural resource stocks, which can be represented by easy-to-interpret economic depreciation measures. Moreover, \cite{Costanza2009} argued that the simplicity of GDP makes it prone to misuse in measuring welfare, and it is therefore necessary to go `beyond GDP' and identify `what matters' to humans. Furthermore, GDP fails to fully capture the contribution of nature to economic activity and human well-being \citep{ouyang2020using}.\par
 In the pioneering work of \cite{costanza1997value}, the global value contributed by ecosystem services to human welfare was explored by using both market and nonmarket valuation methods. Shortly thereafter, many studies established various natural resources accounting (NRA) frameworks, such as natural capital (\cite{macdonald1999applying}), ecosystem services (\cite{polasky2011,kosoy2010payments,paulin2019towards}), gross ecosystem product (\cite{ouyang2013gross,ouyang2020using}), ecological assets (\cite{Obst2016,vavckavruu2019toward}), environmental assets (\cite{bartelmus2014environmental}) and services (\cite{Muradian2010,liu2018meta}). In addition, \cite{hambira2007} developed an NRA framework to integrate water resource management in Botswana into measures of economic well-being. \cite{2007Natural} established an NRA framework to reflect the value of forest resources in India.  There have also been substantial efforts by the United Nations, such as the Millennium Ecosystem Assessment \citep{MEA} and The Economics of Ecosystems and Biodiversity \citep{TEEB}. For recent survey on the development of natural resources accounting, see \cite{zhong2016bibliometric}. \par
\begin{figure}[!ht]
	\centering
	\includegraphics[width=0.85\textwidth]{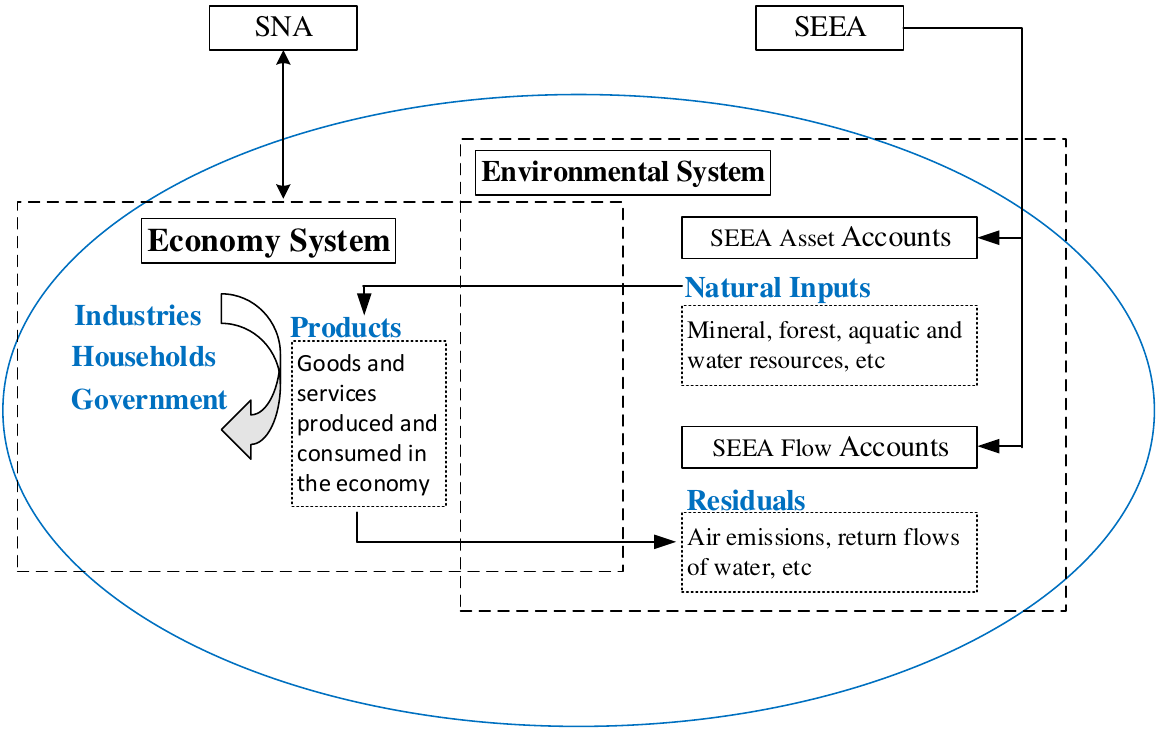}
\caption{The relationship between SNA and SEEA.}
\label{Fig1}
\end{figure}
As a complement to the SNA, the System of Environmental Economic Accounting (SEEA), released by the UN Statistics Division in 1993 \citep{UN1993}, is perhaps one of the most widely used economic accounting frameworks for systematically evaluating environmental resource stocks and flows \citep{Holub1999Some}. To elaborate on its conception and definition, the UN modified the physical and monetary natural resource accounts and the evaluation methods in the SEEA-1993 in 2003. As Figure \ref{Fig1} shows, the SEEA explicitly refers to the natural inputs (minerals, energy, water, forests, etc.) and residuals (air pollutant emissions, return flow of water, etc.) overlooked within the SNA, while the SNA refers only to products (goods, services, etc.). In 2012, the SEEA Central Framework and SEEA Experimental Ecosystem Accounting (SEEA EEA) were developed by the UN, UC and others to develop stock and flow accounts that incorporate natural resources into national accounts in a manner that complies with the SNA (\cite{UN2014}). The SEEA Central Framework includes physical flows of materials and energy, stocks of environmental assets, and environment-related transactions (\cite{LaNottea2020}). Owing to the lack of a requirement that environmental assets provide economic benefits to an economic owner, the scope of environmental assets measured in the SEEA Central Framework is greater than that obtained by following the SNA definition of economic assets (\cite{UN2014}, Paragraph 5.39). The SEEA Central Framework includes resource overexploitation, pollutant emissions, and ecological degradation; however, there is still no consensus on the accounting scope 
 and methods \citep{ouyang2018}. Even the SEEA Central Framework does not mention the concept of environmental liability (\cite{UN2014}). This lack of agreement has limited the potential to assign responsibility for environmental pollution and ecological destruction and, most importantly, has hindered the wider acceptance of liability thinking in reversing natural deterioration.\par

\subsection{Ecological liability, environmental liability and natural resource liability}
As \cite{Polasky2015} rightly pointed out, liability rules are a potentially powerful tool for providing incentives to protect the vital natural capital assets that are essential for maintaining the flow of valuable ecosystem services. The concept of ecological liability (or debt), developed by Western countries in the early 1990s, was introduced to increase awareness of the environmental responsibility embedded in the responsibility for past colonial subjugation. In 1992, Jernel\"{o}v calculated Sweden's liability to future generations and defined ecological liability as "the restoration costs for tech-economic environmental harms and the capital required to pay for recurring repair efforts" (\cite{Warlenius2015}). Another approach to ecological liability was developed by \cite{wang2008} to measure the loss of biodiversity in populations of vertebrate species and the emissions of greenhouse gases and pollutants. \cite{srinivasan2008debt} incorporated the concept of ecological debt and estimated the environmental costs of human activities; the estimations showed dramatic losses and degradation of ecosystem assets and flows. \cite{xie2010} defined ecological liability as the disparity in the distribution of ecological service consumption between the ecological footprint and biocapacity. Recently, \cite{ogilvy2015} proposed an ecological balance sheet that provides a more complete representation of an agricultural entity's total equity and that records where changes to ecological asset accounts cause flows of expenditure. Afterwards, \cite{ogilvy2018} introduced an integrated accounting method to calculate the liabilities due to ecosystem degradation and conducted a scenario analysis to model a pastoral livestock operation in leased rangelands in Australia. Currently, \cite{ouyang2018} have proposed a conceptual accounting framework to compile an ecological asset balance sheet for regional ecosystem assessment. A brief historical overview of ecological liability can be found in \cite{Warlenius2015}. \par

Measuring liability for environmental harm plays important roles in both deterrence and corrective justice \citep{phelps2015environmental}. It increases the financial and nonfinancial burdens of rule-breaking in ways that can inhibit future environmental harm while also compensating victims and securing resources for environmental restoration. In 1998, the United Nations Conference on Trade and Development (\cite{UN1998}) defined environmental liability as an obligation relating to the environmental costs that are incurred by an enterprise and that meet the criteria for recognition as a liability. The guidelines for recognition, measurement, and disclosure of environmental liability provide a basis for the development of more robust monetary accounting. To account for environmental liability, \cite{cardoso2015behind} aimed to identify and assign values to the environmental liabilities due to coal mining at different stages of the coal life cycle in Cesar. Regarding the impact of disclosures, \cite{chen2014} found that the overall level of environmental liabilities consistently decreased over the time frame examined, suggesting that earlier adoption would have made more sense. \cite{friehe2017prevention} explored incentives for accident prevention and cleanup when firms are subject to environmental liability reporting. Recently, \cite{negash2020institutional} investigated the role of institutional forces in shaping corporate accounting and reporting practices for environmental liabilities within the context of South Africa. To a certain extent, the existing research has predominantly focused on the antecedents and economic consequences of corporate environmental performance and related disclosures (\cite{schneider2017environmental}).\par

To the best of our knowledge, the concept of natural resource liability, developed by the National Oceanic and Atmospheric Administration of the United States in 1996, was incorporated into the Oil Pollution Act to implement natural resource liability provisions. \cite{Jones1997} investigated a conceptual framework to measure the damage to natural resources and to determine compensation for protecting natural resources from injury and depletion and illustrated the concept of natural resource liability \footnote{Natural resource liability statutes embody a clear preference for trustees to ensure the restoration of injured resources to their baseline levels and for recoveries of interim losses incurred from the time of the incident until resource recovery to be spent on restoring, rehabilitating, replacing, or acquiring the equivalent of the injured natural resources.}. Shortly afterwards, \cite{2000Economic} advised resource managers to use a variety of economic methods to assess the trade-offs between gains from proposed actions and interim losses from injuries. Empirical evidence has indicated that natural resource liability actions represent a potent tool for federal and state resource managers to address injuries to public resources. To overcome the controversy surrounding valuation methods in regulations, \cite{2018The} proposed a service-to-service approach called habitat equivalency analysis (HEA) to assess the injuries to ecological services and adapted its use to applications in which restoration projects make resource and ecosystem service substitutions. These studies have provided a powerful tool for policymakers and managers to protect public natural resources. \par

In 2013, the concept of a natural resource liability appeared in China and was accompanied by the creation of an NRBS. One of the most critical components in the conceptual framework of the NRBS is natural resource liability accounting. As defined by \cite{shujian2016}, natural resource liability is the current responsibility that economic entities have for the excessive consumption of public resources, which hinders production conditions and economic growth. \cite{ouyang2018} defined natural resource liability as the relevant economic value needed to compensate for the ecological value of natural resources. Moreover, most researchers have agreed with the definition of natural resource liability as the previous business activities, accidents, or anticipated events of the natural resource accounting entity that have resulted in the net loss of natural resources and a negative impact on the environment and ecology \citep{shi2020research}. The natural resource liability account consists of three components: the overexploitation of natural resources, environmental pollution and ecological degradation. \par
Therefore, the definition of natural resource liability used by the United States is totally different from that used by China. The former emphasizes the use of legal liability to enable trustees to recover damages for injuries to public resources. The latter attaches importance to the establishment of natural resources liability accounts, which can help halt and reverse anthropogenic damage to nature. These liability accounts enable the government to identify the cost of environmental pollution, ecological degradation and resource exhaustion, assign responsibility for the cost to the entities responsible, and ensure that the entities shift to a growth pattern that enables coexistence with nature. To date, the incorporation of the value of nature resources into liability rules has resulted in a growing recognition of the need to resolve the challenge of the resilience and sustainability of nature.\par
\subsection{Natural resources balance sheets}
As mentioned above, the compilation of NRBS is still exploratory in China, though this fact is less known to international researchers (\cite{song2019}). Existing literature on NRBSs has mostly focused on specific types of natural resources, including
water resource balance sheets (\cite{jia2017ac,song2018,tian2018,tang2020theory}), forest resource balance sheets (\cite{zhang2019study,lin2021rationality}), mineral resource balance sheets (\cite{li2015balance,ji2016study,geng2019}), and marine resource balance sheets (\cite{he2017,li2018}). Recently, researchers have attempted to conduct a subnational NRBS pilot for regional resource restoration, such as that in Huzhou city \citep{yan2017} and Chengde city \citep{yang2017}. The pilot programs have shown that the NRBS is comprehensive, including 72 tables in total. Therefore, the use of the NRBS is difficult and inconvenient for provincial-level natural resource management. Another practical difficulty is estimating the price of natural resources. Moreover, justifications for the existence of natural resource liability and for the related accounting methods are driving innovative work to generate liability information and assign responsibility to decision makers. As pointed out by \cite{Adams2017}, overall responsibility lies with national governments, and sustainable development goals cannot be achieved without a concerted effort from businesses and other organizations through properly integrated reporting models. \cite{song2019} pointed out that the current research on environmental accounting mainly emphasizes micro-level enterprises, while the NRBS focuses on the macro-level government. A lack of adequate supervision of the government and officials tends to entrench authoritarian regimes, to increase certain types of corruption and to trigger violent conflict in low- and middle-income countries \citep{Ross2015What}. However, existing studies on the NRBS have ignored the liabilities in government supervision, which are associated with property rights regimes for natural resources. \par

Our goal is to resolve the above challenges related to (\romannumeral1) estimating the physical quantity and monetary value of natural resource assets; (\romannumeral2) investigating whether natural resource liability should be recognized; (\romannumeral3) estimating the physical quantity and monetary value of natural resource liabilities; (\romannumeral4) compiling a coherent and compliant balance sheet for natural resources; and (\romannumeral5) presenting a theoretical list of responsibility for liabilities based on natural resource property rights regimes. Finally, the measurement and division of natural resources liabilities are employed to assign responsibility to the government and officials.\par
\section{Methodology}
\subsection{Accounting framework}
\begin{figure}[h]
	\centering
	\includegraphics[width=0.8\textwidth]{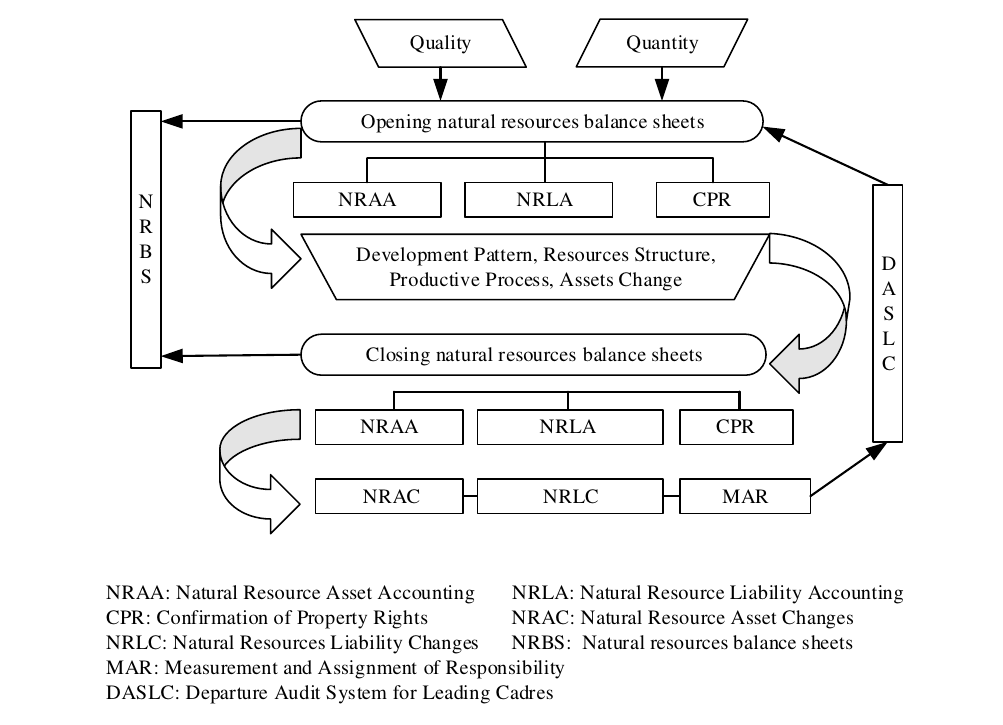}
\caption{Mind map of NRBS compilation.}
\label{Fig2}
\end{figure}
The NRBS accounting framework and its main components are illustrated in Figure \ref{Fig2}. Here, we focus on the elements of the NRBS, a measure that translates natural resources assets held by humans and the degradation of nature into monetary terms. Much of the strength of the balance sheet comes from its intuitive illustration of the financial situation, which can be readily understood by the decision maker. Analogous to standard balance sheets, NRBS use monetary methods and surrogates for monetary methods to calculate natural resource assets and natural resource liabilities and aggregate them into a balance sheet that indicates the relationship between nature and humans. The NRBS are strengthened by the implementation of the Departure Audit System for Leading Cadres (DASLC) and its administrative means for restoring natural conditions. The measurement and assignment of responsibility included in the DASLC enables the government to assign responsibility for liabilities, which indicates that activities to prevent natural resource liabilities can be enforced by public supervision.\par
\subsection{Key accounting concepts}
Natural resource assets are the main contributions of nature, such as land, energy, minerals, water and forests. Natural resource liabilities include resource overexploitation, environmental pollution and ecosystem degradation. Table \ref{tab1} presents a complete list of natural resource assets and liabilities.\par
\begin{table}[htbp]
\caption{NRBS system}
\label{tab1}
\centering
\begin{tabular}{lll}
\toprule[1pt]
Account& Items & Contents \\
\cline{1-3}
\textbf{Asset}&&\\
&Land &Cultivated land, gardens, woodland, grassland, built-up and related  \\
& & \ \ \ \  areas (construction), traffic land, maintenance and restoration of \\
&& \ \ \ \  environmental functions (maintenance) and others\\
&Energy& Oil, Natural gas, Coal\\
&Minerals & Antimony, gold, mercury, lead, molybdenum, zinc, iron,    \\
& & \ \ \ \  sodium salt (NaCl), cement limestone, glass quartzite\\
&Water & Surface water, groundwater, soil water\\
&Forests & Cultivated timber, natural timber\\
&&\\
\textbf{Liability}&&\\
& Resource & Depletion of nonrenewable resources, biodiversity loss\\
&overexploitation&\\
& Environmental & Air pollution, water pollution, solid waste pollution\\
&pollution&\\
& Ecosystem & Climate change, stratospheric ozone-layer depletion, agricultural \\
&degradation&\ \ \ \   intensification and expansion, overfishing, flood prevention and\\
&&\ \ \ \ soil erosion\\
\bottomrule[1pt]
\end{tabular}
\end{table}\par

\textbf{Natural resource asset accounting:} To become a credible index for decision making, the NRBS should preferably be readily calculable from available data. Many types of natural resource data exist from which to develop natural resource asset accounting. In the pioneering work of \cite{UN2014}, resources such as land, energy, minerals, water and forests were considered in the SEEA Environmental Asset Accounts based on a variety of physical measures and monetary methods. However, the issue with using only physical measures is that multiple noncommensurate metrics that lack sufficient credibility for incorporation within resource restoration and management are involved. Focusing on the above valuation methods for natural resources, in this paper, the income capitalization approach is used for different kinds of land, and a credible price is obtained for a variety of energy sources and minerals, which have drastically changing market prices. Data on market prices for water and timber are used, as they are available and accessible. Last, we combine the values for the various natural resources into NRBS asset account and present in {\color{blue}Supplementary Materials}. \par

\textbf{Establishment of the existence of natural resource liability:} There have been concerns and debates about the existence of and justification for natural resource liabilities. According to the stipulations of the SNA 2008 and the SEEA Central Framework, it is inappropriate to confirm the liability status of natural resources with unclear ownership and undefined debtor--creditor relationships. The SEEA Central Framework has established an equilibrium relation in which `the source of the asset=possession of the asset' and has taken no position on the potential to account for nature resource liabilities \citep{song2019}. Moreover, the SEEA Experimental Ecosystem Accounting has taken a relatively negative position on the potential to account for liabilities related to the degradation of ecosystems \citep{ogilvy2018}. Therefore, the functional account setting advocated by the SEEA Central Framework is superior to the liability account included in the NRBS. \par

However, given the standard macroeconomic balance sheet described in the SNA, the accounting equation $\mathrm{Asset = Liability + Owner\ \  equity}$ implies the potential existence of natural resource liabilities. Many researchers have argued that assets and liabilities are a pair of corresponding concepts, which is usually true and is apparent in the SNA \citep{Collis2017}. As mentioned in Section 2.2, natural resource liabilities in the NRBS have prominent Chinese characteristics, as the Chinese government aims to assess the losses from and liability for the exploitation of natural resources and the resultant environmental pollution and ecological destruction \citep{zhiming2014natural}. Furthermore, officials are responsible for maintaining natural conditions, which indicates that the leading cadres in the government should take action to reverse the deterioration of nature, which is enforced by the rules of the DASLC. Therefore, reporting on only the supply and consumption of natural resources cannot indicate whether the use and consumption of natural resources are at reasonable levels \citep{song2019}. Moreover, the environmental and ecological impact of the overexploitation of natural resources has also raised concern about the consequent threat to human well-being \citep{cardoso2015behind}. Most importantly, because the destruction of nature by humans is not fully captured in commercial markets or adequately quantified in economic systems, it is often given too little weight in policy decision making. This neglect may ultimately compromise the sustainability of human civilizations on Earth. This reality has triggered a shift in our attention from moral and emotional obligations to active and effective actions. The existence of natural resource liabilities reveals the attitude of those who are liability towards the protection of nature and provides reliable and useful information for appraising a liable entity's economic development patterns and awareness of environmental and ecological restoration \citep{song2019}. Hence, we completely agree that natural resource liability is not a perfect indictor for measuring the anthropogenic destruction of nature, but it is a good start for mimicking or approximating that liability.\par

\textbf{Natural resource liability accounting:} To become as credible as natural resource assets accounting, natural resource liability accounting must be easily calculable from accessible data and proper monetary methods. \cite{du2018review} decomposed natural resource liability accounting into three components: the overexploitation of natural resources, environmental pollution, and ecological degradation. There are some controversial methodological assumptions related to the assessment of overexploitation. Although the concept of natural resource overexploitation was proposed in \citep{yan2017}, account items and related monetary methods for calculating this type of liability item are still lacking. As mentioned in \cite{wang2020}, China's consumption of energy and of important mineral resources over the past 20 years has been 2-8 times that of the total over the 50 years from 1949 to 1999. Moreover, as pointed out in \cite{lu2020spatial}, there is large spatial variation in the distribution of threatened species across China's provinces, with the biodiversity in several provinces being severely reduced. Therefore, the depletion of nonrenewable resources and biodiversity loss is used as the index to measure natural resource overexploitation. Following \cite{2019Economic}, we use the replacement cost method to estimate the surrogate price of nonrenewable resources. The physical and monetary method of natural resources liabilities are provided in {\color{blue}Supplementary Materials}.\par

As \cite{yang2017} summarized, environmental pollution includes three types of pollution: air, water, and solid waste. Air pollution includes $\mathrm{SO_{2}}$, $\mathrm{NO_{x}}$ and suspended particles. Water pollution is measured by the emission of chemical oxygen demand (COD)
 and ammonia nitrogen. Solid waste includes household and industrial waste. Following \cite{Shen2017}, we use the imputed abatement cost method to estimate the price and cost of pollution. Regarding ecological degradation, there is now an important recognition of the need to restore ecosystems by addressing issues such as greenhouse gas emissions, stratospheric ozone depletion, agricultural intensification and expansion, deforestation, overfishing, and mangrove conversion \citep{srinivasan2008debt}. Our work to develop an ecological degradation account builds on this prior work. In light of the severe problems with flooding and soil erosion in China \citep{liu2020assessment}, we further incorporate government expenditures on flooding and soil erosion prevention into ecological liability. Given the government expenditures, the replacement cost method is used to evaluate the potential liability of ecosystem devastation. \par

\textbf{The principle behind NRBS compilation:} As noted above, the relationship between assets and liabilities is given by the equation `Assets = Liabilities + Owner equity'. Nevertheless, replacing the concept of owner equity with net worth in the NRBS is reasonable. There are several explanations for this change. First, in China, public ownership dominates, which suggests that natural resources are owned by the state (i.e., owned collectively by all individuals). Second, as \cite{song2019} pointed out, the amount of capital invested by the subject entity and the residual earnings cannot be directly calculated so that net worth can be measured only by the balance between natural resource assets and liabilities. Third, natural resources belong to the authorities and are managed by the Chinese Ministry of Natural Resources (MNR) established in 2018. Because of the above concerns, the definition of `an owner' is somewhat unclear and differs with different circumstances. Given the similarity of this situation to that in \citep{ogilvy2018}, we compile the NRBS following the principle that `Net worth $=$ Natural resources assets $-$ Natural resources liabilities'.

\textbf{Assign responsibility for the liability:} The premise underlying the assignment of responsibility for each liability lies with the property rights regimes, as a liability cannot be confirmed without a basis in the laws and regulations issued by National People's Congress of China. The Constitution stipulates that China's natural resources are publicly and collectively owned. \cite{sikor2017property} responded to changes in natural resource governance, stating that modifications to three specific rights---use rights (use of direct benefits, use of indirect benefits), control rights (management, exclusion, transaction, monitoring) and authoritative rights (definition, allocation)---are needed to develop a framework that distinguishes among eight types of property rights. Before 2018, the difference between control rights and authoritative rights was ambiguous. The governments at all levels serve as acting managers and supervisors \citep{song2019}. The collective ownership of natural resources has led to the phenomenon of absentee owners. The state is the virtual subject of all rights. Therefore, the main problems within the natural resources property rights system include unclear property ownership, a severe loss of property rights, deviations from market rule in the exchange of assets and imperfect management systems \citep{ma2015}. These issues hinder the efficiency of utilization in resource management due to a lack of motivation and of supervision. In 2018, the National People's Congress designated the Ministry of Natural Resources (MNR) as trustees of the resources on behalf of the public, which clarifies the ownership of the natural resources and allows the trustees to protect those resources and repair the damages to them. Meanwhile, the National People's Congress has appointed the Ministry of Ecology and Environment (MEE) to restore ecosystems and supervise pollutant emissions. The establishment of two management agencies enables us to assign legal responsibility for potential liabilities.\par
Definitions of natural resource liability consist of four elements: the debtors, the creditors, the repayment period, and expenditures. Since the subject of responsibility in the NRBS is macroscopic, the main creditor for the liability of natural resource overexploitation should be the MNR, while the main creditor for ecosystem degradation and environmental pollution should be the MEE. The debtors may be individuals, companies or organizations that create natural resource liabilities. Given the complexity and urgency of the actual situation, the repayment period remains an unsolved issue. Therefore, the compilation of the NRBS can enable the responsible parties (government and officials) to visualize the costs and act as a reminder the main leading cadres of the need to restore natural conditions. Moreover, it is the responsibility of the supervisors to ensure that activities to remedy natural conditions are enforced through the implementation of the DASLC. \par
\subsection{Compilation steps}
 The compilation of the NRBS involves five steps as follows.
\begin{itemize}
\item Collect the physical quantity and quality data on the different natural resource asset types and natural resource liability types from within the regional area of interest (state, province, city, county).
\item Assess the value of the different natural resource assets based on reliable monetary methods and measure the potential loss from different natural resource liabilities based on widely accepted evaluation methods.
\item Understand the natural resource property rights regimes in the areas of interest, and establish a feasible strategy or mechanism for assigning responsibility for the liability based on the legal and institutional systems at the national and regional levels.
\item Designate the period for which the NRBS is compiled through opening and closing dates and calculate the change in physical quantities and financial quantities over the period.
\item Analyse the gains and losses due to natural resource management and official governance and further present suggestions for policy.
\end{itemize}

\section{Case Study: The NRBS for Shaanxi Province}
\begin{figure}[!h]
	\centering
	\includegraphics[width=0.8\textwidth]{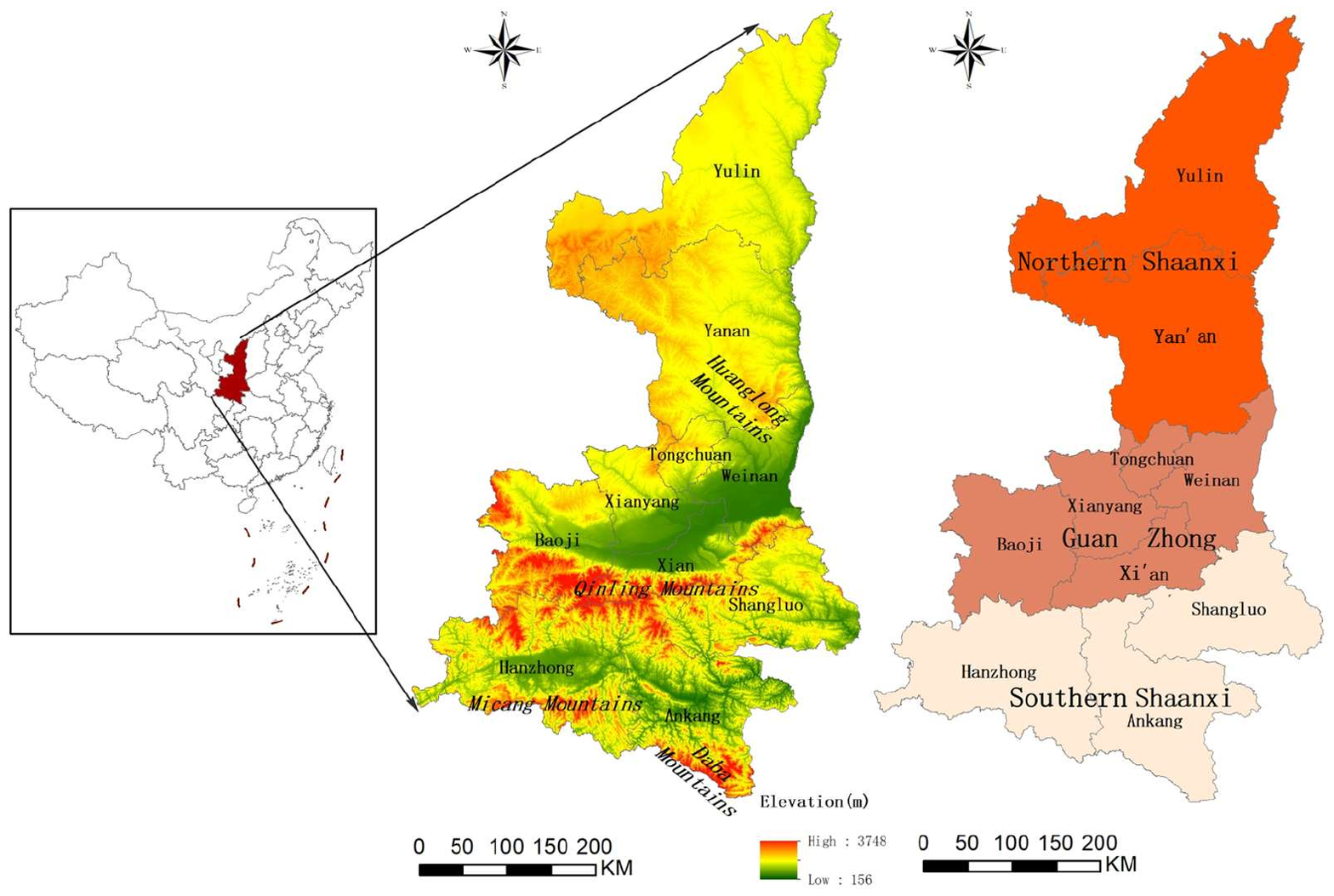}
\caption{Digital elevation map and regional division of Shaanxi Province, China.}
\label{Fig3}
\end{figure}
\subsection{Study area}
A typical representative of the natural environment in East Asia, Shaanxi Province is selected as the study area for several reasons. The primary reason is that Shaanxi Province comprises three distinct natural regions: the mountainous southern region (Southern Shaanxi), the Wei River Valley (Guanzhong), and the northern upland plateau (Northern Shaanxi). Shaanxi Province ($105^{\circ}29^{'}-111^{\circ}15^{'}$E, $31^{\circ}42^{'}-39^{\circ}35^{'}$N) is located in the hinterlands of China (Figure 3) with an area of 210,000 $km^2$ and a population of 39.5 million. The province is bounded by the Qinling Mountains, which display distinct diversity in their natural and geographical features. The pattern of precipitation also varies spatially and temporally across the province, decreasing from south to north and being low in the winter but substantial in the summer. In addition, the temperature, solar radiation and vegetation distributions across the province exhibit spatial heterogeneity. The area is interlaced with mountains, plateaus, basins, and deserts, which have temperate, warm temperate, and northern subtropical climates. \par

Another main reason for selecting this study area is that Shaanxi is an important source of energy fuels (coal, oil, natural gas, etc.) and minerals (iron, molybdenum, copper, etc.) for much of China. The land types include farmlands, grasslands, shrubs, sparse woodlands, and woodlands \citep{liu2019hotspot}. Known as the `central water tower' of China, the Jialingjiang River, Lou River, Wei River and Hanjiang River all originate in the Qinling Mountains. The Qinling Mountains are also rich in mineral resources and have extensive forest coverage. Moreover, Shaanxi provides 22.0 billion $m^{3}$ of water annually to other parts of China. The forest area in Shaanxi covers nearly 8.9 million hectares with reserves of 479.0 million $m^{3}$.\par

A final important reason for choosing Shaanxi Province is that ecological degradation and environmental pollution have led to severe concerns about living conditions and the resultant threat to economic development given the rapid growth in the population and in resource consumption in the province. 
 As noted by \cite{fu2000relationships}, the hilly area in Northern Shaanxi probably has the highest soil erosion rate in the world. The Qinling Mountains are threatened by illegal mining and overcutting. \cite{liu2019hotspot} reported that in early 2000, soil and water loss, deforestation, and desertification had become key concerns for the Shaanxi government. \cite{hou2015gis} criticized Shaanxi for having lost its natural protective and productive functions in China. To reverse the damage and restore its natural conditions, Shaanxi Province has implemented a number of ecological restoration programs (such as the Grain-for-Green Program and the Natural Forest Protection Program), which have improved the status of vegetation in the province, particularly in Northern Shaanxi. Because of these representative features of Shaanxi, compiling the NRBS for this province and exploring the physical and monetary changes in its natural resources are essential for the government and leading cadres in resource management to supervise the ecological environment. 
 \par

\subsection{NRBS Accounting in Shaanxi province}
\begin{table}[!ht]
{\caption{Opening natural resources balance sheets for Shaanxi province---December 31, 2013}\label{tab6}
\scalebox{0.8}{
\begin{tabular*}{\textwidth}{@{\extracolsep{\fill}}llrrcllrr}
\cline{1-9}
\multicolumn{4}{l}{\textbf{Assets}}& &\multicolumn{4}{l}{\textbf{Liabilities}}\\
\cline{1-4}\cline{6-9}
&Accounting&Physical&Monetary&&&Accounting&Physical&Monetary\\
Category&item&quantity&value&&Category&item& quantity&value\\
\cline{1-4}\cline{6-9}
\textbf{Land}&($\times 10^{3}$)& $\mathrm{km}^{2}$
&(billion yuan)&& \textbf{Resource}&($\times 10^{7}$)&t&(billion yuan)\\
&Cultivated land &$39.92$&1054.70&&\textbf{overexploi-}&\textbf{Nonrenewable resource}&&\\
&Gardens&8.26&2892.94&&\textbf{tation}&\textbf{depletion}&&\\
&Woodland&112.03&5652.94&&&Coal&172.47&102.13\\
&Grassland&28.74&1739.60&&&Oil &11.76&59.55\\
&Construction&0.26&11.33&&&($\times 10^{10}$)&$m^{3}$&\\
&Traffic land&2.48&109.19&&&Natural gas&7.03&13.70\\
&Irrigation land &3.10&29.59&&&($\times 10^{7}$)&$t$&\\
\cline{1-4}
\textbf{Energy}&($\times 10^{10}$)&t&&&&Iron&5.03&2.84\\
&Coal&163.97&1393.73&&&Molybdenum&17.09&25.26\\
&Oil &0.34&1706.43&&&Lead&0.04&0.59\\
&($\times 10^{10})$&$m^{3}$&&&&Zinc&0.71&8.75\\
&Natural gas&623.11&1215.08&&&\textbf{Biodiversity loss}&--&--\\
\cline{1-4}\cline{6-9}
\textbf{Minerals}&($\times 10^{3}$)&t&&&\textbf{Environmental}&($\times 10^{7}$)&t&\\
&Antimony&27.55&1.82&&\textbf{pollution}&\textbf{Air pollution}&&\\
&Gold&0.35&82.45&&&$\mathrm{SO_{2}}$&0.81&1.02\\
&Mercury&1.50&0.94&&&$\mathrm{NO_{x}}$&0.76&0.96\\
&($\times 10^{7}$)&t&&&&SP&0.54&0.30\\
&Lead&1.59&22.03&&&\textbf{Water pollution}&&\\
&Molybdenum&1.02&185.62&&&COD&0.52&0.73\\
&Zinc&3.51&43.45&&&Ammonia Nitrogen&0.06&0.10\\
&($\times 10^{10}$)&t&&&&\textbf{Solid waste}&\\
&Iron&0.78&441.68&&&Household waste&4.21&0.11\\
&Sodium salt (NaCl)&885.53&1284.02&&& Solid industrial waste &&\\
&Cement limestone&7.68&7.68&&&Disposal&16.22&1.22\\
&Glass quartzite&0.19&0.22&&&Storage&11.31&0.17\\
\cline{1-4}
\textbf{Water}&($\times 10^{10})$&$m^{3}$&&&&Hazardous industrial waste&&\\
&Surface water&33.15&155.46&&& Disposal&0.12&0.18\\
&Groundwater&11.85&55.59&&&Storage&0.09&0.03\\
\cline{6-9}
&Soil water&145.69&683.30& &\textbf{Ecological}&($\times 10^{7})$&t&\\
\cline{1-4}
\textbf{Forests}&($\times 10^{7})$&$m^{3}$&&&\textbf{degradation}&$\mathrm{CO_{2}}$&265.60&12.47\\
&Cultivated timber&2.36&1.79&&&Stratospheric &&\\
&Natural timber &5.32&4.01&&&ozone layer depletion&--&--\\
\cline{1-4}
& &&&&&($\times 10^{3})$&$\mathrm{hm}^{2}$&\\
& &&&&&Agricultural intensi- &&\\
&&& &&&fication and expansion &4108.22&0.78\\
& &&&&&Flood prevention &&\\
&&& &&&and soil erosion &667.00&0.93\\
&&& &&&Overfishing &--&1.78\\
\cline{6-9}\\
\multicolumn{2}{l}{\textbf{Total Assets:}} &&\textbf{18775.59}&\multicolumn{2}{l}{\textbf{Total Liabilities:}}&&&\textbf{233.60}\\
\cline{1-4}\cline{6-9}
\multicolumn{2}{l}{\textbf{Net Worth:}} &&\textbf{18541.99}&\multicolumn{2}{l}{}&&\\
\cline{1-9}
\end{tabular*}}}
\end{table}\par

\begin{table}[!ht]
{\caption{Closing natural resources balance sheets for Shaanxi province---December 31, 2018}\label{tab7}
\scalebox{0.8}{
\begin{tabular*}{\textwidth}{@{\extracolsep{\fill}}llrrcllrr}
\cline{1-9}
\multicolumn{4}{l}{\textbf{Assets}}& &\multicolumn{4}{l}{\textbf{Liabilities}}\\
\cline{1-4}\cline{6-9}
&Accounting&Physical&Monetary&&&Accounting&Physical&Monetary\\
Category&item&quantity&value&&Category&item& quantity&value\\
\cline{1-4}\cline{6-9}
\textbf{Land}&($\times 10^{3}$)& $\mathrm{km}^{2}$
&(billion yuan)&& \textbf{Resources}&($\times 10^{7}$)&t&(billion yuan)\\
&Cultivated land &$39.77$&1050.68&&\textbf{overexploi-}&\textbf{Nonrenewable resource}&&\\
&Gardens&8.15&2854.25&&\textbf{tation}&\textbf{depletion}&&\\
&Woodland&111.64&5633.29&&&Coal&193.96&127.04\\
&Grassland&28.68&1735.67&&&Oil &7.78&25.97\\
&Construction&0.16&10.27&&&($\times 10^{10}$)&$m^{3}$&\\
&Traffic land&2.64&164.82&&&Natural gas&10.55&20.58\\
&Irrigation land &3.08&29.37&&&($\times 10^{7}$)&$t$&\\
\cline{1-4}
\textbf{Energy}&($\times 10^{10}$)&t&&&&Iron&22.02&11.66\\
&Coal&171.6&1458.98&&&Molybdenum&16.43&28.66\\
&Oil &0.35&1163.97&&&Lead&0.03&0.45\\
&($\times 10^{10})$&$m^{3}$&&&&Zinc&1.10&21.46\\
&Natural gas&959.48&1870.98&&&\textbf{Biodiversity loss}&--&--\\
\cline{1-4}\cline{6-9}
\textbf{Mineral}&($\times 10^{3}$)&t&&&\textbf{Environmental}&($\times 10^{7}$)&t&\\
&Antimony&41.70&1.53&&\textbf{pollution}&\textbf{Air pollution}&&\\
&Gold&0.46&93.75&&&$\mathrm{SO_{2}}$&0.22&0.28\\
&Mercury&1.29&0.67&&&$\mathrm{NO_{x}}$&0.31&0.39\\
&($\times 10^{7}$)&t&&&&SP&0.20&0.11\\
&Lead&2.65&23.62&&&\textbf{Water pollution}&&\\
&Molybdenum&1.40&269.20&&&COD&0.18&0.25\\
&Zinc&4.71&68.42&&&Ammonia Nitrogen&0.02&0.04\\
&($\times 10^{10}$)&t&&&&\textbf{Solid waste}&\\
&Iron&1.10&580.12&&&Household waste&6.38&0.17\\
&Sodium salt (NaCl)&928.60&1346.50&&& Solid industrial waste &&\\
&Cement limestone&7.73&7.73&&&Disposal&66.11&4.96\\
&Glass quartzite&0.19&0.23&&&Storage&15.38&0.23\\
\cline{1-4}
\textbf{Water}&($\times 10^{10})$&$m^{3}$&&&&Hazardous industrial waste&&\\
&Surface water&34.75&163.00&&& Disposal&0.61&0.91\\
&Groundwater&12.50&58.64&&&Storage&0.21&0.06\\
\cline{6-9}
&Soil water&144.53&677.86& &\textbf{Ecological}&($\times 10^{7})$&t&\\
\cline{1-4}
\textbf{Forest}&($\times 10^{7})$&$m^{3}$&&&\textbf{degradation}&$\mathrm{CO_{2}}$&276.17&12.98\\
&Cultivated timber&3.11&1.77&&&Stratospheric &&\\
&Natural timber &5.76&3.28&&&ozone layer depletion&--&--\\
\cline{1-4}
& &&&&&($\times 10^{3})$&$\mathrm{hm}^{2}$&\\
& &&&&&Agricultural intensi- &&\\
&&& &&&fcation and expansion &4092.10&0.78\\
& &&&&&Flood prevention &&\\
&&& &&&and soil erosion &228.80&5.05\\
&&& &&&Overfishing &--&2.98\\
\cline{6-9}\\
\multicolumn{2}{l}{\textbf{Total Assets:}} &&\textbf{19268.60}&\multicolumn{2}{l}{\textbf{Total Liabilities:}}&&&\textbf{265.01}\\
\cline{1-4}\cline{6-9}
\multicolumn{2}{l}{\textbf{Net Worth:}} &&\textbf{19003.59}&\multicolumn{2}{l}{}&&\\
\cline{1-9}
\end{tabular*}}}
\end{table}\par
The opening and closing NRBSs are presented in Tables \ref{tab6} and \ref{tab7} and cover the period from 2013 to 2018. In the opening NRBS from 2013, the natural resource assets in Shaanxi totalled 18541.99 billion yuan, and the liabilities totalled 233.6 billion yuan, indicating that the value of the natural resource assets was greater than the cost of the liabilities. Overall, the various land types contributed 61.2\% of the total value of the asset account. As befits the saying that one-third of the potential value of China is its mineral resources, energy and mineral resources contributed 34\% of the total value. The other main assets include water resources (4.7\%) and forest resources (0.1\%). For the liabilities account, the consumption of energy and minerals reached 212.82 billion yuan, suggesting that the mining industry accounted for 13.3\% of the total value of the regional GDP (1604.52 billion Yuan) in Shaanxi. The second most-important liability component was ecological degradation (6.8\%). Among the items under the heading of ecological degradation, $\mathrm{CO}_{2}$ emissions are the top concern, contributing 12.47 billion yuan. The cost of environmental pollution accounted for 2\% of the total liability. In conclusion, the net worth of Shaanxi was 18541.99 billion in 2013. \par

Table \ref{tab7} lists the closing NRBS for Shaanxi from 2018. As noted above, the assets account is designed to measure the total value of all resources and summarize that value in a single monetary metric. The liabilities account is developed to fully capture the costs of the harm done to nature by humans within a regional area. Tables \ref{tab6} and \ref{tab7} demonstrate that both assets and liabilities increased from 2013 to 2018, suggesting that the NRBS reports on two aspects of change: the benefit humans derive from nature and the damage humans do to nature. According to the equation `Net worth $=$ Natural resource assets $-$ Natural resource liabilities", the net worth of Shaanxi was 18541.99 billion yuan in 2013 and 19003.59 billion yuan in 2018. This suggests an increase of 2.49\% over the course of the five years, implying that the government and resource management officials have potentially made progress. \par

\subsection{Physical and monetary changes between 2013 and 2018}
\begin{table}[!ht]
{\caption{Physical and monetary changes in the natural resource assets account from 2013 to 2018}
\label{tab8}
\scalebox{0.75}{
\begin{tabular*}{\textwidth}{@{\extracolsep{\fill}}llrlrrrrrc}
\cline{1-10}
&& \multicolumn{2}{c}{Physical quantity}&\multicolumn{2}{c}{Constant 2013 price}&\multicolumn{2}{c}{Current price}&\multicolumn{2}{r}{}\\
&& \multicolumn{2}{c}{2013-2018}&\multicolumn{2}{c}{2013-2018}&\multicolumn{2}{c}{2013-2018}&Valuation&Billion yuan\\
\cline{3-4}\cline{5-6}\cline{7-8}\cline{9-10}
Category&Accounting&Level&Quantity&Level&Percent&Level&Percent&Unit&Valuation\\
 of asset&items&change&& change&change&change&change&price&method\\
 \cline{1-10}
\textbf{Land}&&&&&&&&&\\
&Cultivated land &$-0.22\downarrow$&$10^{3} \ \mathrm{km}^{2}$&$-4.02\downarrow$&0.38\%&$-4.02\downarrow$&0.38\%&$26.42\times 10^{7} \mathrm{yuan/km}^{2}$&Income \\
&Gardens&$-0.11\downarrow$&$10^{3} \ \mathrm{km}^{2}$&$-38.69\downarrow$&1.34\%&$-38.69\downarrow$&1.34\%&$350.38\times 10^{7} \mathrm{yuan/km}^{2}$&capitalization \\
&Woodland&$-0.39\downarrow$&$10^{3} \ \mathrm{km}^{2}$&$-19.65\downarrow$&0.35\%&$-19.65\downarrow$&0.35\%&$50.46\times 10^{7} \mathrm{yuan/km}^{2}$&\\
&Grassland&$-0.06\downarrow$&$10^{3} \ \mathrm{km}^{2}$&$-3.93\downarrow$&0.23\%&$-3.93\downarrow$ &0.23\%&$60.53\times 10^{7} \mathrm{yuan/km}^{2}$&\\
&Construction&$-0.10\downarrow$&$10^{3} \ \mathrm{km}^{2}$&$-4.09\downarrow$&36.10\%&$-1.06\downarrow$&9.36\%&$44.01\times 10^{7} \mathrm{yuan/km}^{2}$&Market\\
&Traffic land&$0.16\uparrow$&$10^{3} \ \mathrm{km}^{2}$&$7.04\uparrow$&6.45\%&$55.63\uparrow$&50.95\%&$44.01\times 10^{7} \mathrm{yuan/km}^{2}$&prices\\
&Irrigation land&$-0.02\downarrow$&$10^{3} \ \mathrm{km}^{2}$&$-0.22\downarrow$&0.74\%&$-0.22\downarrow$&0.74\%&$9.55\times 10^{7} \mathrm{yuan/km}^{2}$&\\
\cline{1-10}
\textbf{Energy}&&&&&&&&&\\
&Coal&$7.63\uparrow$&$10^{10}\ t$&$65.25\uparrow$&4.68\%&$65.25\uparrow$&4.68\%&8.50\ yuan/t&Replacement \\
&Oil &$11.76\uparrow$&$10^{7}\ t$&$65.88\uparrow$&3.49\%&$-542.46\downarrow$&31.79\%&5061.64\ yuan/t&costs \\
&Natural gas&$336.37\uparrow$&$10^{10} \ m^{3}$&$655.90\uparrow$&53.98\%&$655.90\uparrow$&53.98\%&1.95\ yuan/$m^{3}$&\\
\cline{1-10}
\textbf{Minerals}&&&&&&&&&\\
&Antimony&$14.15\uparrow$&$10^{3}\ t$&$0.93\uparrow$&51.36\%&$-0.29\downarrow$&15.95\%&$65.95\times10^{3}$\ yuan/t&Market\\
&Gold&$0.11\uparrow$&$10^{3}\ t$&$23.60\uparrow$&28.62\%&$11.30\uparrow$&13.71\%&$21.45\times 10^{7}$ \ yuan/t&prices\\
&Mercury&$-0.21\downarrow$&$10^{3}\ t$&$-0.13\downarrow$&14.07\%&$-0.27\downarrow$&28.72\%&$0.63\times 10^{7}$\ yuan/t&\\
&Lead&$1.06\uparrow$&$10^{7}\ t$&$14.63\uparrow$&66.66\%&$1.59\uparrow$&7.21\%&$1.38\times 10^{4}$\ yuan/t&\\
&Molybdenum&$0.38\uparrow$&$10^{7}\ t$&$56.09\uparrow$&37.25\%&$83.58\uparrow$&45.03\%&$14.76\times 10^{4}$\ yuan/t&\\
&Zinc&$1.20\uparrow$&$10^{7}\ t$&$14.88\uparrow$&32.24\%&$24.97\uparrow$&57.47\%&$1.24\times 10^{4}$\ yuan/t&\\
&Iron&$0.32\uparrow$&$10^{10}\ t$&$180.74\uparrow$&40.92\%&$138.44\uparrow$&31.34\%&564.80\ yuan/t&\\
&Sodium salt (NaCl)&$43.07\uparrow$&$10^{10}\ t$&$62.48\uparrow$&4.87\%&$62.48\uparrow$&4.87\%&1.45 yuan/t&\\
&Cement limestone&$0.05\uparrow$&$10^{10}\ t$&$0.05\uparrow$&0.65\%&$0.05\uparrow$&0.65\%&1.00 yuan/t&\\
&Glass quartzite&$0.002\uparrow$&$10^{9}\ t$&$0.0024\uparrow$&1.09\%&$0.0024\uparrow$&1.09\%&1.20 yuan/t&\\
\cline{1-10}
\textbf{Water}&&&&&&&&&\\
&Surface water&$1.60\uparrow$&$10^{10}\ m^{3}$&$7.54\uparrow$&4.85\%&$7.54\uparrow$&4.85\%&4.85 yuan/$m^{3}$&Market\\
&Groundwater&$0.65\uparrow$&$10^{10}\ m^{3}$&$3.05\uparrow$&5.49\%&$3.05\uparrow$&5.49\%&4.85 yuan/$m^{3}$&prices\\
&Soil water&$-1.16\downarrow$&$10^{10}\ m^{3}$&$-5.44\downarrow$&0.80\%&$-5.44\downarrow$&0.80\%&4.85 yuan/$m^{3}$&\\
\cline{1-10}
\textbf{Forests}&&&&&&&&&\\
&Cultivated timber&$0.75\uparrow$&$10^{7}\ m^{3}$&$0.57\uparrow$&31.78\%&$-0.02\downarrow$&1.11\%&754.00 \ yuan/$m^{3}$&Market\\
&Natural timber&$0.44\uparrow$&$10^{7}\ m^{3}$&$0.33\uparrow$&8.27\%&$-0.73\downarrow$&18.20\%&754.00 \ yuan/$m^{3}$&prices\\
\cline{1-10}
\end{tabular*}}}
\end{table}
As presented in Table \ref{tab8}, the change in the value of the natural resource assets account from 2013 to 2018 can be attributed to changes in physical quantities (constant prices) and current prices (value per unit). First, the majority of the decrease in the value of land resulted from a decrease in the physical quantity of land. Despite the fact that the value per unit of construction land increased from $44.01 \times 10^{7} \mathrm{Yuan}/\mathrm{km}^{2}$ to $62.56 \times 10^{7} \mathrm{Yuan}/\mathrm{km}^{2}$, there was a sharp decrease of $36.10\%$ in construction land between 2013 and 2018. Except for traffic land, the physical quantity and valuation of all the other types of land decreased over the study period. Second, the valuation and quantity of coal and of natural gas increased by 53.98\% and 4.68\%, respectively. Nonetheless, we witnessed an increase in the production of oil, and the value of oil decreased by 31.79\% due to the fall in its unit price. Third, as shown in Table \ref{tab8}, the unit market price severely affected the value of antimony. Except for that of mercury, the valuation of all mineral resources
increased with fluctuations. For water resources, surface water and groundwater increased by 4.85\% and 5.49\%, respectively, while soil water decreased by 0.80\%. Last, the amount of cultivated timber and natural timber produced in Shaanxi increased by 31.78\% and 8.27\%, respectively. However, the value of this timber decreased by 1.11\% and 18.20\% due to changes in the unit price.\par

\begin{table}[!ht]
{\caption{Physical and monetary changes in the natural resource liabilities account from 2013 to 2018}
\label{tab9}
\scalebox{0.7}{
\begin{tabular*}{\textwidth}{@{\extracolsep{\fill}}llrlrrrrrc}
\cline{1-10}
&& \multicolumn{2}{c}{Physical quantity}&\multicolumn{2}{c}{Constant 2013 price}&\multicolumn{2}{c}{Current price}&\multicolumn{2}{c}{}\\
&& \multicolumn{2}{c}{2013-2018}&\multicolumn{2}{c}{2013-2018}&\multicolumn{2}{c}{2013-2018}&Valuation&Billion yuan\\
\cline{3-4}\cline{5-6}\cline{7-8}\cline{9-10}
Category&Accounting&Level&Quantity&Level&Percent&Level&Percent&Unit&Valuation\\
 of liability&items&change&& change&change&change&change&price&method\\
 \cline{1-10}
\textbf{Resource}&&&&&&&&&\\
\textbf{overexploi-}&\textbf{Nonrenewable resource} &&&&&&&&\\
\textbf{tation}&\textbf{depletion}&&&&&&&\\
&Coal&$21.49\uparrow$&$10^{7} \ t$&$12.60\uparrow$&12.34\%&$24.91\uparrow$&24.39\%&586.32 yuan/t&Replacement \\
&Oil&$-3.98\downarrow$&$10^{7} \ t$&$-20.15\downarrow$&33.84\%&$-32.58\downarrow$ &54.71\%&5061.64 yuan/t&costs \\
&Natural gas&$3.52\uparrow$&$10^{10} \ \mathrm{m}^{3}$&$6.86\uparrow$&50.07\%&$6.88\uparrow$&50.22\%&1.95\ yuan/$\mathrm{m}^{3}$&\\
&Iron&$16.99\uparrow$&$10^{7} \ t$&$9.60\uparrow$&337.78\%&$8.82\uparrow$&310.56\%&564.80 yuan/t&\\
&Lead&$-1.24\downarrow$&$10^{4}\ t$&$0.17\downarrow$&28.81\%&$-0.14\downarrow$&23.72\%&$1.38\times 10^{4}$\ yuan/t&\\
&Molybdenum&$-0.66\downarrow$&$10^{7} \ t$&$-0.97\downarrow$&3.84\%&$3.40\uparrow$&13.46\%&$14.76\times 10^{4}$\ yuan/t&\\
&Zinc&$0.39\uparrow$&$10^{7}\ t$&$4.84\uparrow$&55.31\%&$12.71\uparrow$&145.26\%&$1.24\times 10^{4}$\ yuan/t&\\
&\textbf{Biodiversity loss} &$-$&$-$&$-$&$-$&$-$&$-$&$-$&\\
\cline{1-10}
\textbf{Environmental}&&&&&&&&&\\
\textbf{pollution}&\textbf{Air pollution}&&&&&&&& \\
&$\mathrm{SO_{2}}$ &$-0.59\downarrow$&$10^{7}\ t$&$-0.74\downarrow$&72.55\%&$-0.74\downarrow$&72.55\%&1264.00 yuan/t&Imputed \\
&$\mathrm{NO_{x}}$&$-0.45\downarrow$&$10^{7}\ t$&$-0.57\downarrow$&59.38\%&$-0.57\downarrow$&59.38\%&1264.00 yuan/t& abatement\\
&SP&$-0.34\downarrow$&$10^{7}\ t$&$-0.19\downarrow$&63.33\%&$-0.19\downarrow$&63.33\%&550.00Yuan/t&cost \\
&\textbf{Water pollution}&&&&&&&& method\\
&COD&$-0.34\downarrow$&$10^{7}\ t$&$-0.48\downarrow$&65.75\%&$-0.48\downarrow$&65.75\%&1400.00 yuan/t&\\
&Ammonia Nitrogen&$-0.04\downarrow$&$10^{7}\ t$&$-0.06\downarrow$&60.00\%&$-0.06\downarrow$&60.00\%&1750.00 yuan/t&\\
&\textbf{Solid waste}&&&&&&&& \\
&Household waste&$2.17\uparrow$&$10^{7}\ t$&$0.06\uparrow$&54.55\%&$0.06\uparrow$&54.55\%&27.00\ yuan/t&\\
&Industrial solid waste&&&&&&&&\\
&Disposal&$49.89\uparrow$&$10^{7}\ t$&$3.74\uparrow$&306.56\%&$3.74\uparrow$&306.56\%&75.00 yuan/t&\\
&Storage&$4.07\uparrow$&$10^{7}\ t$&$0.06\uparrow$&35.29\%&$0.06\uparrow$&35.29\%&15.00 yuan/t&\\
&Hazardous industrial waste&&&&&&&&\\
&Disposal&$0.49\uparrow$&$10^{7}\ t$&$0.73\uparrow$&465.56\%&$0.73\uparrow$&465.56\%&1500.00 yuan/t&\\
&Storage&$0.12\uparrow$&$10^{7}\ t$&$0.03\uparrow$&132.37\%&$0.03\uparrow$&132.37\%&1.20 yuan/t&\\
\cline{1-10}
\textbf{Ecological}&&&&&&&&&\\
\textbf{degradation}&$\mathrm{CO_{2}}$&$10.57\uparrow$&$10^{7}\ t$&$0.51\uparrow$&4.09\%&$0.51\uparrow$&4.09\%&47.03 yuan/t&Imputed \\
&Stratospheric&&&&&&&&abatement\\
&ozone layer depletion&$-$&$-$&$-$&$-$&$-$&$-$&$-$&costs\\
&Agricultural intensi-&&&&&&&&\\
&fication and expansion&$-16.12\downarrow$&$10^{3}\ \mathrm{hm}^{2}$&$-0.003\downarrow$&0.38\%&$-0.003\downarrow$&0.38\%&190.21 \ yuan/$\mathrm{hm}^{2}$&\\
&Overfishing&$-$&$-$&$1.20\uparrow$&167.48\%&$1.20\uparrow$&167.48\%&$-$&\\
&Flood prevention&&&&&&&&\\
&and soil erosion&$-438.20\downarrow$&$10^{3}\ \mathrm{hm}^{2}$&$4.12\uparrow$&443.01\%&$4.12\uparrow$&443.01\%&$-$&\\
\cline{1-10}
\end{tabular*}}}
\end{table}

Table \ref{tab9} records the change in the value of the natural resource liabilities account for the accounting period from 2013 to 2018. In the overexploitation of natural resources category, replacement costs are used to assess the value of the depletion of nonrenewable resources. For the consumption of nonrenewable resources, the volume of coal, natural gas, iron and zinc consumed dramatically increased, while the volume of oil, lead and molybdenum consumed decreased between 2013 and 2018. Since there are insufficient data to estimate the ecological degradation and environmental pollution caused by the exploitation of nonrenewable resources, which is a major concern in China, the cost of restoring those resources has most certainly been underestimated. Moreover, resource overexploitation dominated the natural resource liabilities account in Shaanxi from 2013 to 2018. Regarding environmental pollution, we know from Table \ref{tab9} that the total tonnage of air pollution and water pollution in Shaanxi decreased by almost two-thirds between 2013 and 2018. Meanwhile, the cost of household waste increased dramatically by 54.55\%. The cost of industrial solid waste disposal and waste storage increased by 306.56\% and 35.29\%, respectively. Industrial solid waste has been the most severe source of environmental pollution. Moreover, it should be noted that hazardous industrial waste disposal and waste storage also rapidly increased by 465.56\% and 132.37\%, warning us to place more emphasis on industrial solid and hazardous waste. Last, the climate change costs from greenhouse gas emissions in Shaanxi increased from $265.60\times 10^{7} t$ to $276.17\times 10^{7} t$. Greenhouse gas emissions have been the main concern related to ecological degradation. Since the quantity of land decreased, the cost of agricultural intensification and expansion also decreased. Overfishing directly affects the biodiversity of rivers and lakes. To reverse the growth in overfishing, the River Chief System was fully implemented in 2017. Although flood prevention and soil erosion have been the most severe issues in the past, the government has increased its investments in different ecological projects to improve soil and water conservation.
\begin{table}[!htbp]
\caption{Changes in the NRBS and GDP in Shaanxi from 2013 to 2018}
\label{tab10}
\centering
\begin{tabular}{lrrrr}
\toprule[1pt]
& \multicolumn{2}{c}{Constant price}& Level change& Percent change \\
\cline{1-5}
Content&2013 & 2018& (billion yuan)& (\%)\\
\cline{1-5}
NRBS&&&&\\
Natural resource assets&18775.59 &19268.60 &493.01 &2.63\%\\
Natural resource liabilities& 233.60& 265.01 &31.41&13.45\%\\
Net worth&18541.99&  19003.59&461.60&2.49\%\\
GDP&1620.55&2394.19&773.64&47.74\%\\
\toprule[1pt]
\end{tabular}
\end{table}

Table \ref{tab10} shows the change in the NRBS and GDP for Shaanxi Province over the 2013--2018 period. By comparing the growth in natural resource assets with that in natural resource liabilities, we observe that although the depletion of nonrenewable resources in the resource overexploitation category is underestimated, the liabilities account still increased by approximately 13.45\%, indicating that we should improve the efficiency of resource utilization rather than simply increase the production of resources. Considering the level change in GDP, it is noteworthy that the ratio of liabilities to GDP in Shaanxi Province has decreased from 14.42\% in 2013 to 11.07\%, indicating that the ratio of growth in the minerals and energy industries to economic growth decreased during the study period. Overall, the NRBS facilitates the analysis of contemporary changes in natural resource assets and liabilities for resource governance.

\subsection{The analytical framework for responsibility for natural resource liabilities}
\begin{figure}[!ht]
	\centering
	\includegraphics[width=1.26\textwidth,angle=90]{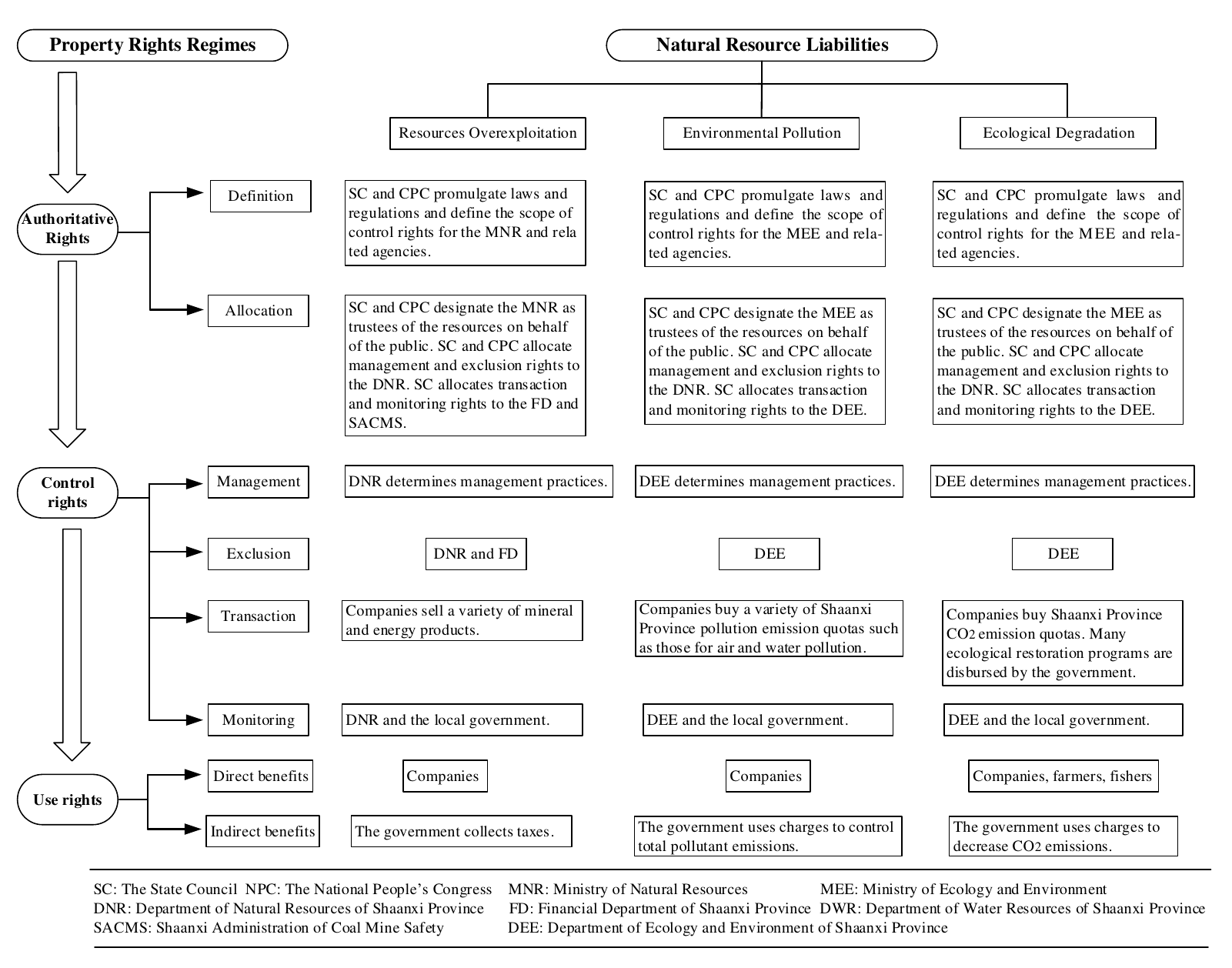}
\caption{Natural resource liabilities with property rights regimes.}
\label{Fig4}
\end{figure}

In practice, property rights over natural resources are often dynamic. It is difficult to analyze natural resource liabilities when property ownership or debtor--creditor relationship is unclear, as they currently are in China \citep{song2019}. Together with the implementation of the DASLC, the goal of measuring and assigning responsibility is to hold the leading cadres accountable for a lack of adequate supervision. To resolve this challenge, we use the direct and indirect beneficiaries of natural resource liabilities as creditors rather than defining a certain creditor. Motivated by the updated conceptual framework for natural resources in \cite{sikor2017property}, the eight types of governmental property rights over natural resource liabilities are developed and arranged in Figure \ref{Fig4}. The three categories of rights are distributed to government departments and ministries and to companies.\par
According to the first column in Figure \ref{Fig4}, since mineral and energy resources are owned by the state, companies possess use rights over all benefits derived from the involved resources by purchasing exploitation rights from the government. In addition, the government auctions off exploitation rights to obtain fiscal revenue. However, once exploitation rights have sold at auction, the government and related officials pay less attention to the mining process and the resultant destruction of the natural environment. In 2017, the Paid Use System for Natural Resources was reformed and re-implemented in China. The government and leading cadres are to guide companies to improve the efficiency of their resource utilization rather than simply increase resource production. Once severe ecological degradation and environmental pollution have occurred due to the  exploitation of mineral and energy resources, the government and relevant officials should use their monitoring rights to manage the related companies and restore the natural environment. \par
The second column in Figure \ref{Fig4} presents the analysis of the power of property rights over environmental pollution. As for pollutant emissions, pilots for the Compensated Use and Trading System for Emission Rights were launched and the full program was implemented in 2014. Emissions quotas for air and water pollutants can be purchased from the government and traded among the various companies. With the efficient management of pollutant emissions, the volume of air and water pollutants decreased from 2013 to 2018, as demonstrated in Table \ref{tab9}, which indicates that emissions quotas are feasible and appropriate mechanisms for managing pollutants. However, it should be noted that emissions quota sales for solid waste are possibly lacking. This is important because, as demonstrated in Table \ref{tab9} and indicated in  Figure \ref{Fig4}, ignoring emissions quotas for solid waste may result in rapid growth in the production of ordinary and hazardous industrial solid waste, and therefore cities may be besieged by waste in the future. Fortunately, the Municipal Solid Waste Classification Policy was implemented in China in 2020. According to Table \ref{Fig4}, the most concerning issue is $\mathrm{CO}_{2}$ emissions. Since the China Carbon Emission Trade Exchange was established on 16 July 2021, the government has used fees to decrease $\mathrm{CO}_{2}$ emissions. Therefore, the government and relevant officials should further introduce a policy for emissions quotas for solid waste, $\mathrm{CO}_{2}$ emissions, especially regular and hazardous industrial solid waste.\par

\section{Conclusions}
\subsection{Main contributions}
This paper provides policy makers with an important snapshot of the financial growing pains experienced by regional governments at specified times by detailing existing natural resource assets and liabilities. As a useful extension of the balance sheet concept, the NRBS enables us to understand the underlying contribution of nature to human well-being and the damage to nature caused by humans. Then, the changes in the physical quantities and monetary values of natural resource assets and liabilities from the beginning period to the closing period can be incorporated into government arrangements and official performance. In addition, we develop a theoretic list of responsibility for liabilities based on various property rights regimes to hold governments and officials responsible for inadequate supervision. The case study shows that the analytical framework for natural resource liabilities is feasible for analyzing and assigning responsibility. \par

\subsection{Policy implications}
Over the past quarter century, China's economy has expanded approximately 10-fold, yet China's GDP orientation cannot last forever due to the related growing pains. The NRBS established in our paper clarifies the danger of ignoring those growing pains, which are a serious detriment to sustainable development. Therefore, the government and related officials should be aware of the destruction of the natural environment, and they have obligations to restore its natural condition and enforce its care through public supervision.\par
A case study has shown the significance of threats to natural resource utilization and the restoration of nature. It is noteworthy that except for traffic land, the quantities of all types of land decreased from 2013 to 2018. Even though this might be only a signal, we should act immediately because the effect of our actions at the macroeconomic level is always lagged \citep{2019Economic}. For natural resource liabilities, the depletion of nonrenewable resources is the herald of a threat, warning us to place more emphasis on the efficiency of resource utilization. Policy makers should adjust their mineral and energy resource management strategies to prevent resource consumption from soaring. In addition, the quantity of $\mathrm{CO}_{2}$ emissions is still growing continuously. In particular, policies such as `Reductions and Limitations in the Coal Industry' and `Carbon Peak and Carbon Neutrality' should be strictly implemented to boost the shift in the economic growth pattern in Shaanxi. Last, the rapid growth in solid waste creates severe environmental pollution. The government and relevant officials should design proper regulations, including emissions quotas for solid waste and price increases for industrial solid waste. Specifically, the Corporate Environmental Protection Rectification Program could be employed to reduce industrial waste-intensive output and promote technological innovation.\par

\subsection{Extensions}
Since the research and practice of NRBS is merely at explanatory stage, we can draw on the experience around world. Based on  "departmental shareholding" of property rights included in SNA and SEEA, researchers may use for reference the experiences of reporting natural resources elements characterized by public property rights in China \citep{song2019}. Moreover, the natural resources of India \citep{2007Natural} and  Australia \citep{ogilvy2018} accounting and reporting practices can also provide an important reference for preparation of NRBS.\par

We would like to discuss some possible topics for future study. First, separating the accounting value of natural resources from the contributions of human labour can be complicated. Second, the limitations in the data and methods lead to imprecise estimations of biodiversity loss and of the ecological cost of mineral and energy exploitation. Furthermore, it need to investigate the relationship between natural resource assets and liabilities under different property rights regimes for other countries. Property rights regimes often vary across countries according to their legislation and practices, designing a mechanism for assigning responsibility for natural resource liabilities deserves to be investigated in the context of different countries.  \par

  \ \ \ \ \ \ \par



\bibliography{wpref}

\begin{thebibliography}{}

\bibitem[Adams, 2017]{Adams2017}
Adams, C.~A. (2017).
\newblock {\em {The Sustainable Development Goals}, ingrated thinking and
  ingrated report}.
\newblock IIRC and ICAS, London.

\bibitem[Adams et~al., 1998]{UN1998}
Adams, R., Coulson, A., Mueller, K., Sturm, A., and Bartel, C. (1998).
\newblock {\em {Accounting and Financial Reporting for Environmental Costs and
  Liabilities}}.
\newblock United Nations Conference on Trade and Development, Geneva.

\bibitem[Allin and Hand, 2017]{Allin2017From}
Allin, P. and Hand, D.~J. (2017).
\newblock From a {System of National Accounts} to a process of national
  wellbeing accounting.
\newblock {\em International Statistical Review}, 85(2):355--370.

\bibitem[Bai et~al., 2014]{bai2014}
Bai, Z., Li, Z., Diao, S., and Yuan, D. (2014).
\newblock Research on urban water price reform in {Shaanxi Province}.
\newblock {\em China Price}, 2014(8):57--60.

\bibitem[Bartelmus, 2014]{bartelmus2014environmental}
Bartelmus, P. (2014).
\newblock Environmental--economic accounting: progress and digression in the
  {SEEA} revisions.
\newblock {\em Review of Income and Wealth}, 60(4):887--904.

\bibitem[Cardoso, 2015]{cardoso2015behind}
Cardoso, A. (2015).
\newblock Behind the life cycle of coal: Socio-environmental liabilities of
  coal mining in {C}esar, {C}olombia.
\newblock {\em Ecological Economics}, 120:71--82.

\bibitem[Chen et~al., 2014]{chen2014}
Chen, J.~C., Cho, C.~H., and Patten, D.~M. (2014).
\newblock Initiating disclosure of environmental liability information: An
  empirical analysis of firm choice.
\newblock {\em Journal of Business Ethics}, 125(4):681--692.

\bibitem[Chen et~al., 2022]{chen2022developing}
Chen, P., Xu, H., Chen, S., Shang, Y., Zhang, B., and Zhao, X. (2022).
\newblock Developing a system framework for {China's} natural resources balance
  sheet from the perspective of sustainable development.
\newblock {\em Frontiers in Environmental Science}, 9:1--9.

\bibitem[Collis et~al., 2017]{Collis2017}
Collis, J., Holt, A., and Hussey, R. (2017).
\newblock {\em The Conceptual Framework for Financial Reporting}.
\newblock International Accounting Standards Board, Cambridge.

\bibitem[Costanza et~al., 1997]{costanza1997value}
Costanza, R., d'{A}rge, R., De~Groot, R., Farber, S., Grasso, M., Hannon, B.,
  Limburg, K., Naeem, S., O'neill, R.~V., Paruelo, J., et~al. (1997).
\newblock The value of the world's ecosystem services and natural capital.
\newblock {\em Nature}, 387(6630):253--260.

\bibitem[Costanza et~al., 2009]{Costanza2009}
Costanza, R., Hart, M., Posner, S., and J, T. (2009).
\newblock {\em {B}eyond {GDP}: the need for new measures of progress}.
\newblock Boston University, Boston.

\bibitem[Du et~al., 2018]{du2018review}
Du, W., Yan, H., and Yan, Y. (2018).
\newblock A review of natural resource asset balance sheets.
\newblock {\em Resources Science}, 40(5):875--887.

\bibitem[Eigenraam and Obst, 2018]{Eigenraam2018}
Eigenraam, M. and Obst, C. (2018).
\newblock Extending the production boundary of the {System of National Accounts
  (SNA)} to classify and account for ecosystem services.
\newblock {\em Ecosystem Health and Sustainability}, 4(11):247--260.

\bibitem[Feng et~al., 2014]{zhiming2014natural}
Feng, Z., Yang, Y., and Li, P. (2014).
\newblock From natural resources accounting to balance-sheet of natural
  resources asset compilation.
\newblock {\em Bulletin of Chinese Academy of Sciences}, 29(4):449--456.

\bibitem[Friehe and Langlais, 2017]{friehe2017prevention}
Friehe, T. and Langlais, E. (2017).
\newblock Prevention and cleanup of dynamic harm under environmental liability.
\newblock {\em Journal of Environmental Economics and Management}, 83:107--120.

\bibitem[Fu et~al., 2000]{fu2000relationships}
Fu, B., Chen, L., Ma, K., Zhou, H., and Wang, J. (2000).
\newblock The relationships between land use and soil conditions in the hilly
  area of the loess plateau in northern {Shaanxi, China}.
\newblock {\em Catena}, 39(1):69--78.

\bibitem[Geng et~al., 2019]{geng2019}
Geng, J., Lyu, X., Shi, J., and Liu, S. (2019).
\newblock Discussion on the compilation and application of energy and mineral
  resources balance sheet.
\newblock {\em China Land Resources Economy}, 32(2):4--14.

\bibitem[Gundimeda et~al., 2007]{2007Natural}
Gundimeda, H., Sukhdev, P., Sinha, R.~K., and Sanyal, S. (2007).
\newblock Natural resource accounting for indian states -- illustrating the
  case of forest resources.
\newblock {\em Ecological Economics}, 61(4):635--649.

\bibitem[Hambira, 2007]{hambira2007}
Hambira, W.~L. (2007).
\newblock Natural resources accounting: A tool for water resources management
  in {Botswana}.
\newblock {\em Physics and Chemistry of the Earth, Parts A/B/C},
  32(15-18):1310--1314.

\bibitem[Hartwick, 1990]{Hartwick1990Natural}
Hartwick, J.~M. (1990).
\newblock Natural resources, national accounting and economic depreciation.
\newblock {\em Journal of Public Economics}, 43(3):291--304.

\bibitem[He et~al., 2017]{he2017}
He, Y., Yue, X., Yang, M., Hu, J., and Li, Y. (2017).
\newblock The balance sheet of {China's} national marine resource assets.
\newblock {\em Marine Development and Management}, 34(10):72--76.

\bibitem[Holub et~al., 1999]{Holub1999Some}
Holub, H.~W., Tappeiner, G., and Tappeiner, U. (1999).
\newblock Some remarks on the `{S}ystem of {I}ntegrated {E}nvironmental and
  {E}conomic {A}ccounting' of the {U}nited {N}ations.
\newblock {\em Ecological Economics}, 29(3):329--336.

\bibitem[Hou et~al., 2015]{hou2015gis}
Hou, K., Li, X., and Zhang, J. (2015).
\newblock {GIS analysis of changes in ecological vulnerability using a SPCA
  model in the Loess plateau of Northern Shaanxi, China}.
\newblock {\em International Journal of Environmental Research and Public
  Health}, 12(4):4292--4305.

\bibitem[Ji and Liu, 2016]{ji2016study}
Ji, X. and Liu, Y. (2016).
\newblock Study on the framework of mineral resource balance sheet.
\newblock {\em China Population, Resources and Environment}, 26(3):100--108.

\bibitem[Jia et~al., 2017]{jia2017ac}
Jia, L., Gan, H., Wang, L., and Qin, C. (2017).
\newblock Accounting methodology of the balance sheet for water resources.
\newblock {\em Journal of Hydraulic Engineer}, 48(11):1324--1333.

\bibitem[Jones, 2000]{2000Economic}
Jones, C.~A. (2000).
\newblock Economic valuation of resource injuries in natural resource liability
  suits.
\newblock {\em Journal of Water Resources Planning \& Management},
  126(6):358--365.

\bibitem[Jones and DiPinto, 2018]{2018The}
Jones, C.~A. and DiPinto, L. (2018).
\newblock The role of ecosystem services in {USA} natural resource liability
  litigation.
\newblock {\em Ecosystem Services}, 29:333--351.

\bibitem[Jones and Pease, 1997]{Jones1997}
Jones, C.~A. and Pease, K.~A. (1997).
\newblock Restoration-based compensation measures in natural resource liability
  statutes.
\newblock {\em Contemporary Economic Policy}, 15(4):111--122.

\bibitem[Kosoy and Corbera, 2010]{kosoy2010payments}
Kosoy, N. and Corbera, E. (2010).
\newblock Payments for ecosystem services as commodity fetishism.
\newblock {\em Ecological Economics}, 69(6):1228--1236.

\bibitem[La~Notte and Rhodesb, 2020]{LaNottea2020}
La~Notte, A. and Rhodesb, C. (2020).
\newblock The theoretical frameworks behind integrated environmental,
  ecosystem, and economic accounting systems and their classifications.
\newblock {\em Environmental Impact Assessment Review}, 80(2020):1--10.

\bibitem[Li and Zhang, 2015]{li2015balance}
Li, H. and Zhang, X. (2015).
\newblock Balance sheet compilation of mineral resources based on {SEEA}
  framework.
\newblock {\em Resources \& Industries}, 17(5):60--65.

\bibitem[Li et~al., 2018]{li2018}
Li, Y., Wei, X., Liu, D., Dai, J., and Chen, K.~y. (2018).
\newblock Study on the formulation framework of marine resources balance sheet
  for marine management.
\newblock {\em Marine Science Bulletin}, 37(3):27--34.

\bibitem[Liang et~al., 2015]{Liang2015}
Liang, D., Liu, T., and Li, Y. (2015).
\newblock Comparative study of {BRICS'} {$\mathrm{CO_2}$} emission cost and its
  influential factors based on {LMDI} model.
\newblock {\em Resources Science}, 37(12):2319--2329.

\bibitem[Lin et~al., 2021]{lin2021rationality}
Lin, M., Jin, L., and Tian, G. (2021).
\newblock The rationality of compiling a forest resource balance sheet.
\newblock {\em Forest Products Journal}, 71(4):309--321.

\bibitem[Liu et~al., 2020]{liu2020assessment}
Liu, B., Xie, Y., Li, Z., Liang, Y., Zhang, W., Fu, S., Yin, S., Wei, X.,
  Zhang, K., Wang, Z., et~al. (2020).
\newblock The assessment of soil loss by water erosion in {China}.
\newblock {\em International Soil and Water Conservation Research},
  8(4):430--439.

\bibitem[Liu et~al., 2019]{liu2019hotspot}
Liu, L., Zhang, H., Gao, Y., Zhu, W., Liu, X., and Xu, Q. (2019).
\newblock Hotspot identification and interaction analyses of the provisioning
  of multiple ecosystem services: Case study of {Shaanxi Province, China}.
\newblock {\em Ecological Indicators}, 107(105566):1--13.

\bibitem[Liu and Kontoleon, 2018]{liu2018meta}
Liu, Z. and Kontoleon, A. (2018).
\newblock Meta-analysis of livelihood impacts of payments for environmental
  services programmes in developing countries.
\newblock {\em Ecological Economics}, 149:48--61.

\bibitem[Long and Ji, 2019]{2019Economic}
Long, X. and Ji, X. (2019).
\newblock Economic growth quality, environmental sustainability, and social
  welfare in {C}hina - provincial assessment based on {Genuine Progress
  Indicator (GPI)}.
\newblock {\em Ecological Economics}, 159:157--176.

\bibitem[Lu et~al., 2020]{lu2020spatial}
Lu, Y., Yang, Y., Sun, B., Yuan, J., Yu, M., Stenseth, N.~C., Bullock, J.~M.,
  and Obersteiner, M. (2020).
\newblock Spatial variation in biodiversity loss across {China} under multiple
  environmental stressors.
\newblock {\em Science Advances}, 6(47):eabd0952.

\bibitem[Ma and Liu, 2015]{ma2015}
Ma, Y. and Liu, Q. (2015).
\newblock The strategic consideration on natural resources property right
  system construction in {China}.
\newblock {\em Bulletin of Chinese Academy of Sciences}, 30(4):503--508.

\bibitem[MacDonald et~al., 1999]{macdonald1999applying}
MacDonald, D.~V., Hanley, N., and Moffatt, I. (1999).
\newblock Applying the concept of natural capital criticality to regional
  resource management.
\newblock {\em Ecological Economics}, 29(1):73--87.

\bibitem[Muradian et~al., 2010]{Muradian2010}
Muradian, R., Corbera, E., Pascual, U., Kosoy, N., and May, P.~H. (2010).
\newblock Reconciling theory and practice: An alternative conceptual framework
  for understanding payments for environmental services.
\newblock {\em Ecological Economics}, 69(6):1202--1208.

\bibitem[Negash and Lemma, 2020]{negash2020institutional}
Negash, M. and Lemma, T.~T. (2020).
\newblock Institutional pressures and the accounting and reporting of
  environmental liabilities.
\newblock {\em Business Strategy and the Environment}, 2020:1--20.

\bibitem[Obst et~al., 2016]{Obst2016}
Obst, C., Hein, L., and Edens, B. (2016).
\newblock National accounting and the valuation of ecosystem assets and their
  services.
\newblock {\em Environmental \& Resource Economics}, 64(1):1--23.

\bibitem[Ogilvy, 2015]{ogilvy2015}
Ogilvy, S. (2015).
\newblock Developing the ecological balance sheet for agricultural
  sustainability.
\newblock {\em Sustainability Accounting, Management and Policy Journal},
  6(2):110--137.

\bibitem[Ogilvy et~al., 2018]{ogilvy2018}
Ogilvy, S., Burritt, R., Walsh, D., Obst, C., Meadows, P., Muradzikwa, P., and
  Eigenraam, M. (2018).
\newblock Accounting for liabilities related to ecosystem degradation.
\newblock {\em Ecosystem Health and Sustainability}, 4(11):261--276.

\bibitem[Opaluch, 2020]{opaluch2020liability}
Opaluch, J.~J. (2020).
\newblock Liability for natural resource damages from oil spills: a survey.
\newblock {\em International Review of Environmental and Resource Economics},
  14(1):37--111.

\bibitem[Ouyang and Jin, 2018]{ouyang2018}
Ouyang, Z. and Jin, L. (2018).
\newblock {\em Developing Gross Ecosystem Product and Ecological Asset
  Accounting for Eco-compensation}.
\newblock Science Press, Beijing.

\bibitem[Ouyang et~al., 2020]{ouyang2020using}
Ouyang, Z., Song, C., Zheng, H., Polasky, S., Xiao, Y., Bateman, I.~J., Liu,
  J., Ruckelshaus, M., Shi, F., Xiao, Y., et~al. (2020).
\newblock Using gross ecosystem product {(GEP)} to value nature in decision
  making.
\newblock {\em Proceedings of the National Academy of Sciences},
  117(25):14593--14601.

\bibitem[Ouyang et~al., 2013]{ouyang2013gross}
Ouyang, Z., Zhu, C., Yang, G., Xu, W., Zheng, H., Zhang, Y., and Xiao, Y.
  (2013).
\newblock Gross ecosystem product concept accounting framework and case study.
\newblock {\em Acta Ecologica Sinica}, 33(21):6747--6761.

\bibitem[Pan et~al., 2022]{pan2022liability}
Pan, X., Song, M., Wang, Y., Shen, Z., Song, J., Xie, P., and Pan, X. (2022).
\newblock Liability accounting of natural resource assets from the perspective
  of input slack—an analysis based on the energy resource in 282
  prefecture-level cities in china.
\newblock {\em Resources Policy}, 78:102867.

\bibitem[Paulin et~al., 2019]{paulin2019towards}
Paulin, M., Remme, R., van~der Hoek, D., de~Knegt, B., Koopman, K., Breure, A.,
  Rutgers, M., and de~Nijs, T. (2019).
\newblock Towards nationally harmonized mapping and quantification of ecosystem
  services.
\newblock {\em Science of the Total Environment}, 703(2020):1--12.

\bibitem[Phelps et~al., 2015]{phelps2015environmental}
Phelps, J., Jones, C.~A., Pendergrass, J., and G\'{o}mez-Baggethun, E. (2015).
\newblock Environmental liability: A missing use for ecosystem services
  valuation.
\newblock {\em Proceedings of the National Academy of Sciences},
  112(39):5379--5379.

\bibitem[Polasky et~al., 2015]{Polasky2015}
Polasky, S., Guerry, A.~D., Lubchenco, J., and Ruckelshaus, M. (2015).
\newblock Reply to {Phelps} et al: Liability rules provide incentives to
  protect natural capital.
\newblock {\em Proceedings of the National Academy of Sciences},
  112(39):5380--5380.

\bibitem[Polasky et~al., 2011]{polasky2011}
Polasky, S., Nelson, E., Pennington, D., and Johnson, K.~A. (2011).
\newblock The impact of land-use change on ecosystem services, biodiversity and
  returns to landowners: a case study in the state of {Minnesota}.
\newblock {\em Environmental and Resource Economics}, 48(2):219--242.

\bibitem[Ross, 2015]{Ross2015What}
Ross, M.~L. (2015).
\newblock What have we learned about the resource curse?
\newblock {\em Annual Review of Political Science}, 18(1):239--259.

\bibitem[Rounsevell et~al., 2020]{rounsevell2020biodiversity}
Rounsevell, M.~D., Harfoot, M., Harrison, P.~A., Newbold, T., Gregory, R.~D.,
  and Mace, G.~M. (2020).
\newblock A biodiversity target based on species extinctions.
\newblock {\em Science}, 368(6496):1193--1195.

\bibitem[Schneider et~al., 2017]{schneider2017environmental}
Schneider, T., Michelon, G., and Maier, M.~S. (2017).
\newblock Environmental liabilities and diversity in practice under
  international financial reporting standards.
\newblock {\em Accounting, Auditing \& Accountability Journal}, 30(2):378--403.

\bibitem[Shan et~al., 2017]{shan2017data}
Shan, Y., Huang, D., Zheng, H., Ou, J., Li, Y., Meng, J., Mi, Z., Liu, Z., and
  Zhang, Q. (2017).
\newblock Data descriptor: {China CO2} emission accounts 1997-2015.
\newblock {\em Scientific Data}, 1(170201):1--14.

\bibitem[Shan et~al., 2020]{shan2020china}
Shan, Y., Huang, Q., Guan, D., and Hubacek, K. (2020).
\newblock {China CO2} emission accounts 2016--2017.
\newblock {\em Scientific Data}, 7(1):1--9.

\bibitem[Shan et~al., 2016]{shan2016new}
Shan, Y., Liu, J., Liu, Z., Xu, X., Shao, S., Wang, P., and Guan, D. (2016).
\newblock New provincial {CO2 emission inventories in China }based on apparent
  energy consumption data and updated emission factors.
\newblock {\em Applied Energy}, 184:742--750.

\bibitem[Shen et~al., 2017]{Shen2017}
Shen, X., Wang, G., and Huang, X. (2017).
\newblock Green {GDP}: Accounting and spatio-temporal pattern in {China} from
  1997 to 2013.
\newblock {\em Journal of Natural Resources}, 32(10):1639--1650.

\bibitem[Shi and Wang, 2020]{shi2020research}
Shi, D. and Wang, J. (2020).
\newblock Research status, literature review and improvement direction of the
  natural resource balance sheet.
\newblock {\em China Population, Resources and Environment}, 30(1):1--11.

\bibitem[Sikor et~al., 2017]{sikor2017property}
Sikor, T., He, J., and Lestrelin, G. (2017).
\newblock Property rights regimes and natural resources: a conceptual analysis
  revisited.
\newblock {\em World Development}, 93:337--349.

\bibitem[Song et~al., 2020]{song2020influences}
Song, M., Ma, X., Shang, Y., and Zhao, X. (2020).
\newblock Influences of land resource assets on economic growth and fluctuation
  in china.
\newblock {\em Resources Policy}, 68:101779.

\bibitem[Song et~al., 2019]{song2019}
Song, M., Shuai, Z., Jing, W., and Shuhong, W. (2019).
\newblock China's natural resources balance sheet from the perspective of
  government oversight: Based on the analysis of governance and accounting
  attributes.
\newblock {\em Journal of Environmental Management}, 248:1--16.

\bibitem[Song et~al., 2018]{song2018}
Song, X., Chen, Y., Yan, H., Yang, Y., and Feng, Z. (2018).
\newblock Initial research into an accounting framework for a water resource
  balance sheet.
\newblock {\em Resources Science}, 40(5):899--907.

\bibitem[Srinivasan et~al., 2008]{srinivasan2008debt}
Srinivasan, U.~T., Carey, S.~P., Hallstein, E., Higgins, P.~A., Kerr, A.~C.,
  Koteen, L.~E., Smith, A.~B., Watson, R., Harte, J., and Norgaard, R.~B.
  (2008).
\newblock The debt of nations and the distribution of ecological impacts from
  human activities.
\newblock {\em Proceedings of the National Academy of Sciences},
  105(5):1768--1773.

\bibitem[Tang et~al., 2020]{tang2020theory}
Tang, Y., Wang, Y., Yang, Q., Hwang, E., and He, Q. (2020).
\newblock Theory and practice of water resources balance sheet compilation in
  the {Yangtze River Basin}.
\newblock {\em Journal of Coastal Research}, 105(SI):71--75.

\bibitem[Tegtmeier and Duffy, 2004]{tegtmeier2004external}
Tegtmeier, E.~M. and Duffy, M.~D. (2004).
\newblock External costs of agricultural production in the {United States}.
\newblock {\em International Journal of Agricultural Sustainability},
  2(1):1--20.

\bibitem[Tian et~al., 2018]{tian2018}
Tian, J., Jiang, T., Shi, H., Zang, N., Li, X., Guo, Y., Mi, X., Zhang, H., and
  Chen, L. (2018).
\newblock Water resource accounting: a case study of water resources stock and
  change table at the {Beilun District,Ningbo City}.
\newblock {\em China Population, Resources and Environment}, 28(9):167--176.

\bibitem[Tschofen et~al., 2019]{tschofen2019fine}
Tschofen, P., Azevedo, I.~L., and Muller, N.~Z. (2019).
\newblock Fine particulate matter damages and value added in the {US} economy.
\newblock {\em Proceedings of the National Academy of Sciences},
  116(40):19857--19862.

\bibitem[UN, 1993]{UN1993}
UN (1993).
\newblock {\em { System of Integrated Environmental and Economic Accounting}}.
\newblock United Nations, New York.

\bibitem[UN et~al., 2014]{UN2014}
UN, EC, FAO, IMF, OECD, and Bank, W. (2014).
\newblock {\em {System of Environmental-Economic Accounting 2012: Central
  Framework}}.
\newblock United Nations, New York.

\bibitem[{United Nations}, 2005]{MEA}
{United Nations} (2005).
\newblock {Millennium Ecosystem Assessment}.
\newblock \url{https://www.millenniumassessment.org/}.

\bibitem[{United Nations}, 2007]{TEEB}
{United Nations} (2007).
\newblock {The Economics of Ecosystems and Biodiversity}.
\newblock \url{http://teebweb.org/}.

\bibitem[Va\v{c}k\'{a}\v{r}{\aa} and Grammatikopoulou,
  2019]{vavckavruu2019toward}
Va\v{c}k\'{a}\v{r}{\aa}, D. and Grammatikopoulou, I. (2019).
\newblock Toward development of ecosystem asset accounts at the national level.
\newblock {\em Ecosystem Health and Sustainability}, 5(1):36--46.

\bibitem[Wang and Gao, 2020]{wang2020}
Wang, A. and Gao, X. (2020).
\newblock {China's} energy and important mineral resources demand perspective.
\newblock {\em Bulletin of Chinese Academy of Sciences}, 35(3):338--344.

\bibitem[Wang and Zhang, 2008]{wang2008}
Wang, Q.~H. and Zhang, Z.~Q. (2008).
\newblock Earth's ecological debt and humankind's sustainable development
  challenges: analysis of {WWF's Living Planet Report} 2006.
\newblock {\em Acta Ecologica Sinica}, 28(5):2424--2429.

\bibitem[Warlenius et~al., 2015]{Warlenius2015}
Warlenius, R., Pierce, G., and Ramasar, V. (2015).
\newblock Reversing the arrow of arrears: The concept of ``ecological debt''
  and its value for environmental justice.
\newblock {\em Global Environmental Change}, 30(2015):21--30.

\bibitem[Xiang and Zheng, 2016]{shujian2016}
Xiang, S. and Zheng, R. (2016).
\newblock Research on the liabilities of natural resources in the balance sheet
  of natural resources.
\newblock {\em Statistical Research}, 33(12):74--83.

\bibitem[Xie et~al., 2010]{xie2010}
Xie, G., Cao, S., Lu, C., Xiao, Y., and Zhang, Y. (2010).
\newblock Human's consumption of ecosystem services and ecological debt in
  {China}.
\newblock {\em Journal of Natural Resources}, 25(1):43--51.

\bibitem[Yan et~al., 2017]{yan2017}
Yan, H., Feng, Z., Yang, Y., Pan, T., Jiang, D., Song, X., Ma, G., and Liu, W.
  (2017).
\newblock First report of the national natural resources balance sheet for
  {Huzhou City and Anji County}.
\newblock {\em Resources Science}, 39(9):1634--1645.

\bibitem[Yan et~al., 2018]{2018yan}
Yan, Y., Chen, Y., Song, X., Yan, H., and Feng, Z. (2018).
\newblock Compilation of a water resource balance sheet for {Huzhou City}.
\newblock {\em Resources Science}, 40(5):908--918.

\bibitem[Yang et~al., 2020]{yang2020}
Yang, S., Tan, Z., and Wang, S. (2020).
\newblock On the method logic and system framework construction of natural
  resource asset and liability accounting in {China}.
\newblock {\em Management World}, 36(11):132--142.

\bibitem[Yang et~al., 2017]{yang2017}
Yang, Y., Feng, Z., Yan, H., Pan, T., Jiang, D., Song, X., Ma, G., and Liu, W.
  (2017).
\newblock The pattern of compilation of the natural resources balance sheet for
  {Chengde City}.
\newblock {\em Resources Science}, 39(9):1646--1657.

\bibitem[Zhang and Li, 2019]{zhang2019study}
Zhang, W. and Li, C. (2019).
\newblock Study on the accounting system of forest resources balance sheet.
\newblock {\em Journal of Natural Resources}, 34(6):1245--1258.

\bibitem[Zhong et~al., 2016]{zhong2016bibliometric}
Zhong, S., Geng, Y., Liu, W., Gao, C., and Chen, W. (2016).
\newblock A bibliometric review on natural resource accounting during
  1995--2014.
\newblock {\em Journal of Cleaner Production}, 139:122--132.

\bibitem[Zhu and Du, 2017]{Zhu2017}
Zhu, D. and Du, T. (2017).
\newblock Accounting method and result analysis of cultivated land resource
  asset in {C}hina.
\newblock {\em China Land Sciences}, 31(10):23--31.

\bibitem[Zhu et~al., 2021]{zhu2021natural}
Zhu, D.-L., Duan, W.-J., Zhang, H., and Du, T. (2021).
\newblock Natural resource balance sheet compilation: a land resource asset
  accounting case.
\newblock {\em Journal of Chinese Governance}, 6(4):515--536.

\end{thebibliography}
\section{Supplemental Materials}
\textbf{Data resources:} We collect data on the physical quantities of the natural resource assets and liabilities in Shaanxi Province from 2013 to 2018. Most of the asset data are obtained from the Annual Shaanxi Provincial Land and Resources Bulletin, the Shaanxi Statistical Yearbook, and the Shaanxi Water Resources Bulletin on official websites. The liability data are acquired from the China Energy Statistical Yearbook, the Shaanxi Provincial Land and Resources Bulletin, the Shaanxi Statistical Yearbook, the China Urban Construction Statistical Yearbook, the Shaanxi Provincial Environmental Bulletin, the Carbon Emission Accounts and Datasets, and the China Statistical Yearbook on the Environment released by the government and scientific institutions. A detailed list of the physical quantities and unit prices of the natural resource assets and liabilities is reported in the Tables \ref{tab2}-\ref{tab5}.\par
\textbf{Monetary methods:} We assessed the monetary value of the natural resource assets and liabilities in Shaanxi Province using a variety of methods. For regulated natural resources, the market price is used. For the value of land, due to the public and collective nature of
ownership, we had access only to information about rents for cultivated land, gardens, woodland and grassland from the Chinese land transfer websites. The web crawler tool Bazhuayu (downloaded from \url{www.bazhuayu.com/}) was used to acquire the total rent data available on the land transfer websites. We deleted observations with missing data and cleaned the data to obtain the average rent for land in Shaanxi. The interest rate and the earning period are provided by \cite{Zhu2017} and the government, respectively. Then, the income capitalization approach is used to estimate the land prices. For the value of liabilities, the methods and related estimation processes  in different disciplines are complex and complicated. Therefore, we apply a variety of market and nonmarket valuations published in the international literature to convert the provision of liabilities into monetary estimates of the anthropogenic damage to nature.

\subsection{Indicators and Methodologies of NRBS}
\begin{table}[!ht]
{\caption{Items description and monetary methodologies of natural resources assets}
\label{tab2}
\scalebox{0.85}{
\begin{tabular*}{\textwidth}{@{\extracolsep{\fill}}lllllllllll}
\cline{1-11}
Accounting& Item&\multicolumn{7}{c}{Monetary Method}&\multicolumn{2}{l}{Data Source}\\
\cline{1-11}\\
\textbf{Land}&Cultivated land &\multicolumn{7}{l}{Income capitalization approach (ICA) is used for valuing }&\multicolumn{2}{l}{\textbf{Physical Quantity}}\\
 &Garden&\multicolumn{7}{l}{the cultivated land at provincial scale.The calculation}&\multicolumn{2}{l}{Shaanxi Provincial Land and}\\
 &Woodland&\multicolumn{7}{l}{formula is $V=\frac{P}{R}\big(1-\frac{1}{(1+R)^{N}}\big)$, where $V$ is the present}&\multicolumn{2}{l}{Resources Bulletin in 2013}\\
 &Grassland&\multicolumn{7}{l}{value of the land, $P$ total rents of land, $R$ is the rate of}&\multicolumn{2}{l}{Shaanxi Provincial Land and}\\
 &&\multicolumn{7}{l}{interest to land, $N$ is the year of earning period. }&\multicolumn{2}{l}{Resources Bulletin in 2018}\\
 &&\multicolumn{7}{l}{According to \cite{Zhu2017}, the price of unit $\mathrm{hm}^{2}$}&\multicolumn{2}{l}{\textbf{Rents }}\\
 &&\multicolumn{7}{l}{for Shaanxi Province is 158889.75 Yuan in 2016.}&\multicolumn{2}{l}{http://www.tuliu.com/}\\
&&\multicolumn{7}{l}{The above ICA method is also used for valuing}&\multicolumn{2}{l}{http://www.jutubao.com/}\\
&&\multicolumn{7}{l}{garden, woodland and grassland at provincial scale.}&\multicolumn{2}{l}{https://www.cnhnb.com/p/tudi/}\\
&&\multicolumn{7}{l}{The total rents of garden, woodland and grassland }&\multicolumn{2}{l}{\textbf{Interest Rate }}\\
&&\multicolumn{7}{l}{are obtained from the right websites. The average of}&\multicolumn{2}{l}{\cite{Zhu2017}}\\
&&\multicolumn{7}{l}{interest rate is $5\%$. According to the Land Admini-}&\multicolumn{2}{l}{\textbf{Earning Period:}}\\
&&\multicolumn{7}{l}{stration Law of the People's Republic of China, $N$ is 30.}&\multicolumn{2}{l}{http://www.npc.gov.cn/}\\
\cline{3-9}\\
&Construction&\multicolumn{7}{l}{Data on the land amount of construction, traffic and}&\multicolumn{2}{l}{\textbf{Land Acquisition Price}}\\
&Traffic land&\multicolumn{7}{l}{maintenance are obtained from Shaanxi Provincial Land}&\multicolumn{2}{l}{http://zrzyt.shaanxi.gov.cn}\\
&Maintenance&\multicolumn{7}{l}{and Resources Bulletin. The average price of unit $\mathrm{hm}^{2}$}&\multicolumn{2}{l}{/info/1150/2780.htm}\\
&&\multicolumn{7}{l}{for land acquisition is  $44.01\times 10^{7} \mathrm{yuan/km}^{2}$ given}&\multicolumn{2}{l}{http://zrzyt.shaanxi.gov.cn}\\
&&\multicolumn{7}{l}{ by Department of Natural Resources of Shaanxi Province.}&\multicolumn{2}{l}{/info/1150/40632.htm}\\
\cline{3-9}\\
&Irrigation land&\multicolumn{7}{l}{The price of irrigation land valued by \cite{costanza1997value}}&\multicolumn{2}{l}{}\\
&&\multicolumn{7}{l}{ (\$ 14785/ha) is used. Data on the change of irrigation land}&\multicolumn{2}{l}{}\\
&&\multicolumn{7}{l}{ are collected in China Forestry Statistical Yearbook. }&\multicolumn{2}{l}{}\\
\cline{1-11}\\
\textbf{Energy}&Oil &\multicolumn{7}{l}{Replacement costs method is used for valuing energy }&\multicolumn{2}{l}{\textbf{Physical Quantity}}\\
&Natural gas &\multicolumn{7}{l}{resources such as oil, natural gas and coal. Since}&\multicolumn{2}{l}{Shaanxi Statistical}\\
&Coal&\multicolumn{7}{l}{the historical price of energy resource in Shaanxi is}&\multicolumn{2}{l}{Yearbook 2013}\\
& &\multicolumn{7}{l}{unavailable. We refer to the historical price in the US.}&\multicolumn{2}{l}{Shaanxi Statistical}\\
&&\multicolumn{7}{l}{The US annual average price of the oil in 2013 and 2018}&\multicolumn{2}{l}{Yearbook 2018}\\
&&\multicolumn{7}{l}{are \$100.95 and \$57.77/barrel. The replacement cost}&\multicolumn{2}{l}{\textbf{Price reference}}\\
&&\multicolumn{7}{l}{of gas between 2013 and 2018 in Shaanxi is 1.95 Yuan/$m^{3}$.}&\multicolumn{2}{l}{https://www.statista.com/}\\
&&\multicolumn{7}{l}{The replacement costs of coal in 2013 and 2018 are \$84}&\multicolumn{2}{l}{https://inflationdata.com/}\\
&&\multicolumn{7}{l}{and \$107/t.}&\multicolumn{2}{l}{}\\
\cline{1-11}\\
\textbf{Mineral}&Sodimu Salt NaCl &\multicolumn{7}{l}{ The market value method is used for valuing mineral }&\multicolumn{2}{l}{\textbf{Physical Quantity}}\\
&Rock Gold&\multicolumn{7}{l}{resources such as salt, gold, molybdenum and so on.}&\multicolumn{2}{l}{Shaanxi Statistical}\\
&Placer Gold&\multicolumn{7}{l}{The price of mineral resources in China is unavailable.}&\multicolumn{2}{l}{Yearbook 2013}\\
&Associated God&\multicolumn{7}{l}{unavailable. We refer to the historical price in the United }&\multicolumn{2}{l}{Shaanxi Statistical}\\
&Molybdenum&\multicolumn{7}{l}{States Geological Survey (USGS). The annual price of }&\multicolumn{2}{l}{Yearbook 2018}\\
&Lead&\multicolumn{7}{l}{salt, gold, molybdenum, lead, zinc, mercury, antimony, }&\multicolumn{2}{l}{\textbf{Price reference}}\\
&Zinc&\multicolumn{7}{l}{limestone, quartzite and iron are provided in the National  }&\multicolumn{2}{l}{https://www.usgs.gov/}\\
&Mercury&\multicolumn{7}{l}{Minerals Information Center of USGS. Data on the amount  }&\multicolumn{2}{l}{centers/nmic/commodity-}\\
&Antimony&\multicolumn{7}{l}{of mineral resources in Shaanxi province are obtained from }&\multicolumn{2}{l}{statistics-and-information}\\
&Cement Limestone&\multicolumn{7}{l}{Shaanxi Statistical Yearbook 2013 and 2018. Data on the}&\multicolumn{2}{l}{}\\
&Glass Quartzite&\multicolumn{7}{l}{other mineral resources are not recorded. The total}&\multicolumn{2}{l}{}\\
&Iron&\multicolumn{7}{l}{monetary value of mineral is therefore underestimated. }&\multicolumn{2}{l}{}\\
\cline{1-11}
\end{tabular*}}}
\end{table}

\begin{table}[!ht]
{\caption{Items description and monetary methodologies of natural resources assets (continued)}
\label{tab3}
\scalebox{0.8}{
\begin{tabular*}{\textwidth}{@{\extracolsep{\fill}}lllllllllll}
\cline{1-11}
Accounting& Item&\multicolumn{7}{c}{Monetary Method}&\multicolumn{2}{l}{Data Source}\\
\cline{1-11}\\
\textbf{Water}&Surface water&\multicolumn{7}{l}{According to \cite{2018yan}, water asset increases=}&\multicolumn{2}{l}{\textbf{Physical Quantity}}\\
&Groundwater &\multicolumn{7}{l}{rainfall+water inflows+socio-economic water return+others;}&\multicolumn{2}{l}{Shaanxi Water Resources}\\
&Soil water&\multicolumn{7}{l}{water asset decreases=water utility+water outflows+others;}&\multicolumn{2}{l}{Bulletin 2013}\\
& &\multicolumn{7}{l}{closing stocks=opening stocks +water asset increases}&\multicolumn{2}{l}{Shaanxi Water Resources}\\
&&\multicolumn{7}{l}{$-$water asset decreases. The unit price of water per $m^{3}$}&\multicolumn{2}{l}{Bulletin 2018}\\
&&\multicolumn{7}{l}{estimated by Shaanxi Administration for Commodity Prices }&\multicolumn{2}{l}{http://slt.shaanxi.gov.cn/}\\
&&\multicolumn{7}{l}{is applied for valuing water price in Shaanxi province}&\multicolumn{2}{l}{\textbf{Price reference}}\\
&&\multicolumn{7}{l}{The average price between 2013 and 2018 is 4.69 Yuan/$m^{3}$.}&\multicolumn{2}{l}{\cite{bai2014}}\\
\cline{1-11}\\
\textbf{Forest}&Cultivated timber&\multicolumn{7}{l}{Market value method is used for valuing both natural }&\multicolumn{2}{l}{\textbf{Physical Quantity}}\\
&Natural timber &\multicolumn{7}{l}{and cultivated timber. According to China Forestry}&\multicolumn{2}{l}{China Forestry Statistical }\\
&&\multicolumn{7}{l}{Statistical Yearbook, the unit price of timer in 2013 and }&\multicolumn{2}{l}{Yearbook 2013}\\
& &\multicolumn{7}{l}{2018 are 754 Yuan/$m^3$ and 569 Yuan/$m^{3}$. Data on}&\multicolumn{2}{l}{China Forestry Statistical}\\
&&\multicolumn{7}{l}{non-timer forest are unavailable. The total forest asset }&\multicolumn{2}{l}{Yearbook 2018}\\
&&\multicolumn{7}{l}{is therefore underestimated.}&\multicolumn{2}{l}{}\\
\cline{1-11}
\end{tabular*}}}
\end{table}
\begin{table}[!ht]
{\caption{Items description and monetary methodologies of natural resources liabilities}
\label{tab4}
\scalebox{0.8}{
\begin{tabular*}{\textwidth}{@{\extracolsep{\fill}}lllllllllll}
\cline{1-11}
Accounting& Item&\multicolumn{7}{c}{Monetary Method}&\multicolumn{2}{l}{Data Source}\\
\cline{1-11}\\
\textbf{Resources}&Depletion of&\multicolumn{7}{l}{According to \cite{2019Economic}, the replacement cost }&\multicolumn{2}{l}{\textbf{Physical Quantity}}\\
\textbf{Over-exploitation}&non-renewable &\multicolumn{7}{l}{method is employed. Since the price of energy and }&\multicolumn{2}{l}{China Energy}\\
&Biodiversity loss&\multicolumn{7}{l}{mineral resource is included, the detailed estimation }&\multicolumn{2}{l}{Statistical Yearbook}\\
& &\multicolumn{7}{l}{is omitted here. Data on the consumption of energy resources}&\multicolumn{2}{l}{Shaanxi Provincial Land }\\
&&\multicolumn{7}{l}{are collected in China Energy Statistical Yearbook. Data}&\multicolumn{2}{l}{and Resources Bulletin}\\
&&\multicolumn{7}{l}{on consumption of mineral resource are not reported in }&\multicolumn{2}{l}{}\\
&&\multicolumn{7}{l}{any official statistics.  We convert the consumption of }&\multicolumn{2}{l}{}\\
&&\multicolumn{7}{l}{mineral resource into equivalent its production. Data on }&\multicolumn{2}{l}{}\\
&&\multicolumn{7}{l}{the production of mineral resource are obtained from}&\multicolumn{2}{l}{}\\
&&\multicolumn{7}{l}{Shaanxi Provincial Land and Resources Bulletin. Data}&\multicolumn{2}{l}{}\\
&&\multicolumn{7}{l}{on the biodiversity loss of Shaanxi are unavailable.}&\multicolumn{2}{l}{}\\
&&\multicolumn{7}{l}{We have to ignore the loss of biodiversity.}&\multicolumn{2}{l}{}\\
\cline{1-11}\\
\textbf{Environmental}&Air pollution&\multicolumn{7}{l}{The air pollution is rooted in the emission of $\mathrm{SO_{2}}$, $\mathrm{NO_{x}}$}&\multicolumn{2}{l}{\textbf{Physical Quantity}}\\
\textbf{Pollution}&&\multicolumn{7}{l}{and Suspended Particles (SP). The unit environmental}&\multicolumn{2}{l}{Shaanxi Statistical}\\
\textbf{}&&\multicolumn{7}{l}{cost of $\mathrm{SO_{2}}$, $\mathrm{NO_{x}}$ and SP estimated by \cite{Shen2017}}&\multicolumn{2}{l}{Yearbook 2013}\\
&&\multicolumn{7}{l}{is 1264 Yuan/t, 1264 Yuan/t and 550 Yuan/t, respectively. }&\multicolumn{2}{l}{Shaanxi Statistical}\\
&Water pollution  &\multicolumn{7}{l}{The water pollution includes the emission of COD and}&\multicolumn{2}{l}{Yearbook 2018}\\
&&\multicolumn{7}{l}{Ammonia Nitrogen. The unit cost of COD and Ammonia}&\multicolumn{2}{l}{China Urban}\\
&&\multicolumn{7}{l}{Nitrogen given by \cite{Shen2017} is 1400 Yuan/t}&\multicolumn{2}{l}{Construction Statistical}\\
&&\multicolumn{7}{l}{and 1750 Yuan/t, respectively. }&\multicolumn{2}{l}{Yearbook 2013.}\\
&Solid waste&\multicolumn{7}{l}{Types of solid waste are household and industrial. The }&\multicolumn{2}{l}{China Urban}\\
& &\multicolumn{7}{l}{cost of the industrial solid waste includes general indust-}&\multicolumn{2}{l}{Construction Statistical}\\
&&\multicolumn{7}{l}{rial solid waste disposal and storage as well as hazardous}&\multicolumn{2}{l}{Yearbook 2018.}\\
& &\multicolumn{7}{l}{industrial solid waste disposal and storage. According to }&\multicolumn{2}{l}{Shaanxi Provincial}\\
&&\multicolumn{7}{l}{\cite{2019Economic}, the unit cost of general industrial }&\multicolumn{2}{l}{Environmental}\\
&&\multicolumn{7}{l}{solid waste disposal and storage is 75 Yuan/t and}&\multicolumn{2}{l}{Bulletin 2013}\\
&&\multicolumn{7}{l}{15 Yuan/t. The unit cost of hazardous industrial solid }&\multicolumn{2}{l}{Shaanxi Provincial}\\
&&\multicolumn{7}{l}{waste disposal and storage is 1500 Yuan/t and 300 Yuan/t.}&\multicolumn{2}{l}{Environmental}\\
&&\multicolumn{7}{l}{The unit cost of household solid waste is 27 Yuan/t.}&\multicolumn{2}{l}{Bulletin 2018}\\
&&\multicolumn{7}{l}{All the data of environmental pollution are obtained from }&\multicolumn{2}{l}{}\\
&&\multicolumn{7}{l}{Shaanxi Statistical Yearbook except for the amount of}&\multicolumn{2}{l}{}\\
&&\multicolumn{7}{l}{household solid waste. The latter data are recorded in}&\multicolumn{2}{l}{}\\
&&\multicolumn{7}{l}{China Urban Construction Statistical Yearbook. Imputed}&\multicolumn{2}{l}{}\\
&&\multicolumn{7}{l}{Abatement Cost Method is used in the above items. }&\multicolumn{2}{l}{}\\
\cline{1-11}
\end{tabular*}}}
\end{table}
\begin{table}[!ht]
{\caption{Items description and monetary methodologies of natural resources liabilities (continued)}
\label{tab5}
\scalebox{0.8}{
\begin{tabular*}{\textwidth}{@{\extracolsep{\fill}}lllllllllll}
\cline{1-11}
Accounting& Item&\multicolumn{7}{c}{Monetary Method}&\multicolumn{2}{l}{Data Source}\\
\cline{1-11}
\textbf{Ecological}&Climate change&\multicolumn{7}{l}{The driver of
 climate change is the emission of}&\multicolumn{2}{l}{\textbf{Physical Quantity}}\\
\textbf{Degradation}&&\multicolumn{7}{l}{greenhouse gas (GHG)(\cite{srinivasan2008debt}).}&\multicolumn{2}{l}{CEADs: Carbon Emission}\\
\textbf{}&&\multicolumn{7}{l}{The average cost of GHG emissions between 1992 and}&\multicolumn{2}{l}{ Accounts and Datasets}\\
\textbf{}&&\multicolumn{7}{l}{2012 estimated by \cite{Liang2015} is 47.03 Yuan/t.}&\multicolumn{2}{l}{https://www.ceads.net/}\\
\textbf{}&&\multicolumn{7}{l}{Data on GHG are obtained from the CEADs database}&\multicolumn{2}{l}{}\\
\textbf{}&&\multicolumn{7}{l}{provided by \cite{shan2017data,shan2020china,shan2016new}.}&\multicolumn{2}{l}{}\\
\textbf{}&Stratospheric ozone- &\multicolumn{7}{l}{The ozone depleting substances (ODS) is rooted in the }&\multicolumn{2}{l}{}\\
&layer depletion &\multicolumn{7}{l}{emission of Chlorofluorocarbon. The related data}&\multicolumn{2}{l}{}\\
&&\multicolumn{7}{l}{are unavailable for Shaanxi Province during 2013-2018.}&\multicolumn{2}{l}{}\\
&Agricultural intensifi- &\multicolumn{7}{l}{Following \cite{srinivasan2008debt}, the agricultural}&\multicolumn{2}{l}{Shaanxi Statistical}\\
&cation and expansion &\multicolumn{7}{l}{intensification and expansion is measured by the}&\multicolumn{2}{l}{Yearbook 2018}\\
&&\multicolumn{7}{l}{external costs of crop production. The united external}&\multicolumn{2}{l}{}\\
&&\multicolumn{7}{l}{cost of crop estimated by \cite{tegtmeier2004external} is }&\multicolumn{2}{l}{}\\
& &\multicolumn{7}{l}{applied. The annual external costs of crop production for}&\multicolumn{2}{l}{}\\
& &\multicolumn{7}{l}{water,soil,air,biodiversity,human-health are 11.50,85.86,}&\multicolumn{2}{l}{}\\
& &\multicolumn{7}{l}{10.85, 43.35, 38.64 Yuan/$\mathrm{hm}^{2}$. Data on the area of crops}&\multicolumn{2}{l}{}\\
& &\multicolumn{7}{l}{are available in the Chapter of Total Sown Areas of Major}&\multicolumn{2}{l}{}\\
& &\multicolumn{7}{l}{Farm of Shaanxi Statistical Yearbook.}&\multicolumn{2}{l}{}\\
&Overfishing &\multicolumn{7}{l}{The value of overfishing is mainly about the gross output}&\multicolumn{2}{l}{Shaanxi Statistical}\\
& &\multicolumn{7}{l}{value of fishery. Data on the value of fishery}&\multicolumn{2}{l}{Yearbook 2018}\\
& &\multicolumn{7}{l}{are available in the Chapter of Gross Output Value of}&\multicolumn{2}{l}{}\\
& &\multicolumn{7}{l}{Fishery of Shaanxi Statistical Yearbook.}&\multicolumn{2}{l}{}\\
&Flood prevention &\multicolumn{7}{l}{Cost of flood prevention and soil erosion is estimated}&\multicolumn{2}{l}{China Statistical Yearbook}\\
&and soil erosion &\multicolumn{7}{l}{by ecological projects the investment of soil and water}&\multicolumn{2}{l}{on Environment 2018}\\
& &\multicolumn{7}{l}{conservation. Since only economic costs are considered,}&\multicolumn{2}{l}{}\\
& &\multicolumn{7}{l}{this term is underestimated. The replacement cost}&\multicolumn{2}{l}{}\\
& &\multicolumn{7}{l}{method is employed in the above items.}&\multicolumn{2}{l}{}\\
\cline{1-11}
\end{tabular*}}}
\end{table}

\subsection{The data of natural resources assets and liabilities}
 The physical quantities and unit prices of the natural resource assets and liabilities from 2013 to 2018 for Shaanxi province is attached in six tables in supplemental materials. The support data of the NRBS involves six tables as follows. Supplementary data related to this article can be found online.

\begin{table}[htbp]
\caption{Data collection}
\label{tabx}
\centering
\scalebox{1.2}{
\begin{tabular}{ll}
\toprule[1pt]
 Items & Files \\
\toprule[1pt]
Assets content in Chinese &Table 1 assets2013-2018 (Chinese).xlsx  \\
 Assets content in English& Table 2 assets2013-2018 (English).xlsx\\
Liabilities content in Chinese& Table 3 liability2013-2018(Chinese).xlsx\\
Liabilities content in Chinese& Table 4 liability2013-2018(English).xlsx\\
Changes in  Assets& Table 5 assets2013-2018 (English).xlsx \\
Changes in  Liabilities & Table 6 liability2013-2018(English).xlsx\\
\toprule[1pt]
\end{tabular}}
\end{table}


\end{document}